**Paper**

# Electromagnetically Driven Thermal Dissipation Scaling in Plasma Centrifuges for Mass Separation

**Drue P. Hood-McFadden[1], Shreyas Kotla[1] and Thomas C. Underwood[1,2,*]**

[1]Department of Aerospace Engineering & Engineering Mechanics, The University of Texas at Austin, Austin, Texas 78712 United States

[2]Texas Materials Institute, The University of Texas at Austin, Austin, Texas 78712, USA
*Author to whom any correspondence should be addressed.

**E-mail:** thomas.underwood@utexas.edu

**Keywords:** magnetohydrodynamics, centrifuge, rotating plasmas, J×B, chemical separation, thermal transport, scaling laws, mass separation

---

**Abstract**
Electromagnetically driven centrifuges (EMDCs) rotate fluids using the Lorentz force, but their separative performance is limited by thermal dissipation that is coupled to electromagnetic forcing. This follows because local centrifugal strength is characterized by $\lambda = mV_\theta^2/2k_BT$, which compares directed kinetic energy that drives species separation to thermal energy that smooths concentration gradients and counteracts separation. In this work, we develop a reduced two-temperature magnetohydrodynamic model to determine how radial geometry, current density, and magnetic field strength control the coupled evolution of rotation and heating that dictates λ. The model is benchmarked against argon velocity, temperature, and pressure measurements spanning current density up to 15 kA/m$^2$, magnetic field up to 0.57 T, feed pressures of 0.5-3 Torr, and annulus sizes of 1-5 cm. A separate Ar/Kr experiment provides qualitative evidence of mass-dependent radial compositional redistribution and is used to evaluate whether the model captures the expected separation mechanism under low-field conditions. The results show that volumetric Lorentz forcing sustains elevated λ, and therefore greater local compositional shifts throughout a larger radial portion of the fluid volume than wall-bounded shear centrifuges, despite producing a lower peak λ. Simulations of a $^{40}$Ar/$^{36}$Ar isotopic mixture demonstrate that, when electromagnetic force and geometry are jointly optimized, EMDCs can approach or match the separative performance of shear-driven centrifuges operating near material speed limits while requiring lower area-averaged values of λ, and can sustain greater radial compositional shifts than the SDC reference over much of the annulus. These results challenge assertions that viscous dissipation constrains weakly ionized plasma centrifuges to $\lambda < 1$, and indicate that enhanced radial mass separation can be achieved by broadening the radial region over which λ remains elevated rather than maximizing its peak or area-averaged value.

---

## 1. Introduction

Mass separation in centrifuges is governed by the balance between centrifugal forcing, which pushes species radially outward, concentration gradients that establish the mass distribution with the lighter species preferentially inward, and thermal motion, which remixes the flow. The device separation factor, $\alpha_{\text{device}}$, captures the magnitude of the radial compositional shift across a separation device and scales exponentially with rotational kinetic energy relative to thermal energy [1] (Fig. 1B). Shear-driven centrifuges (SDCs) exploit this scaling using high wall speeds of roughly 350-700 m/s and near-ambient gas temperatures, but their achievable rotation is limited ultimately by rotor material strength and vibrational imbalances [2, 3]. Modern carbon-fiber SDCs therefore operate near practical wall-speed limits of approximately 700 m/s. For example, in a coaxial geometry at 700 m/s, 300 K, an outer wall radius of $R_2 = 6$ cm, and aspect ratio $R_2/R_1 =$ 3, the ideal isothermal rigid-body derived separation factor ($\alpha_{device,RB}$) for $^{235}UF_6/^{238}UF_6$ is approximately $\alpha_{device,RB} \sim 1.3$ [3].

More generally, the rotational energy required to obtain a given $\alpha_{device}$ depends on the mass difference, $\Delta m$, between species within a flow. It also depends on the relative strength of centrifugal forces and thermal mixing, which scales with $V_\theta^2/T$ where T is the fluid temperature and $V_\theta$ is the rotation speed (Fig. 1B). These dependencies are combined in the centrifugal parameter, $\lambda = mV_\theta^2/2k_BT$, which compares the directed rotational energy of a species with its stochastic thermal energy. The device separation factor is then given by $\alpha_{device} = \exp(\int_{R_1}^{R_2}(2\Delta\lambda/r)\,dr)$, where $\Delta\lambda = \Delta m V_\theta^2/(2k_BT)$ is the difference in the centrifugal parameter between two species that equilibrate to a common rotation speed and temperature. Thus, increasing either $\Delta m$ or $V_\theta^2/T$ increases separation, allowing a given $\alpha_{device}$ to be achieved at lower rotation speeds as $\Delta m$ increases. This scaling is reflected in industrial liquid-solid and liquid-liquid SDCs, which operate at wall speeds of order $10^1$-$10^2$ m/s, with high-speed chemical-processing separators approaching 200 m/s [4]. The reduced velocity requirement for larger $\Delta m$ extends the viability of centrifuge-based separation beyond isotope enrichment toward the high-throughput processing of mixed feedstocks, including critical-material recovery from complex mixtures [5,6]. Furthermore, because $\alpha_{device}$ depends on the radial integral of $\Delta\lambda/r$, rather than only its peak value, both the spatial distribution of $V_\theta^2/T$ and the integration distance provide additional routes to increase separation.

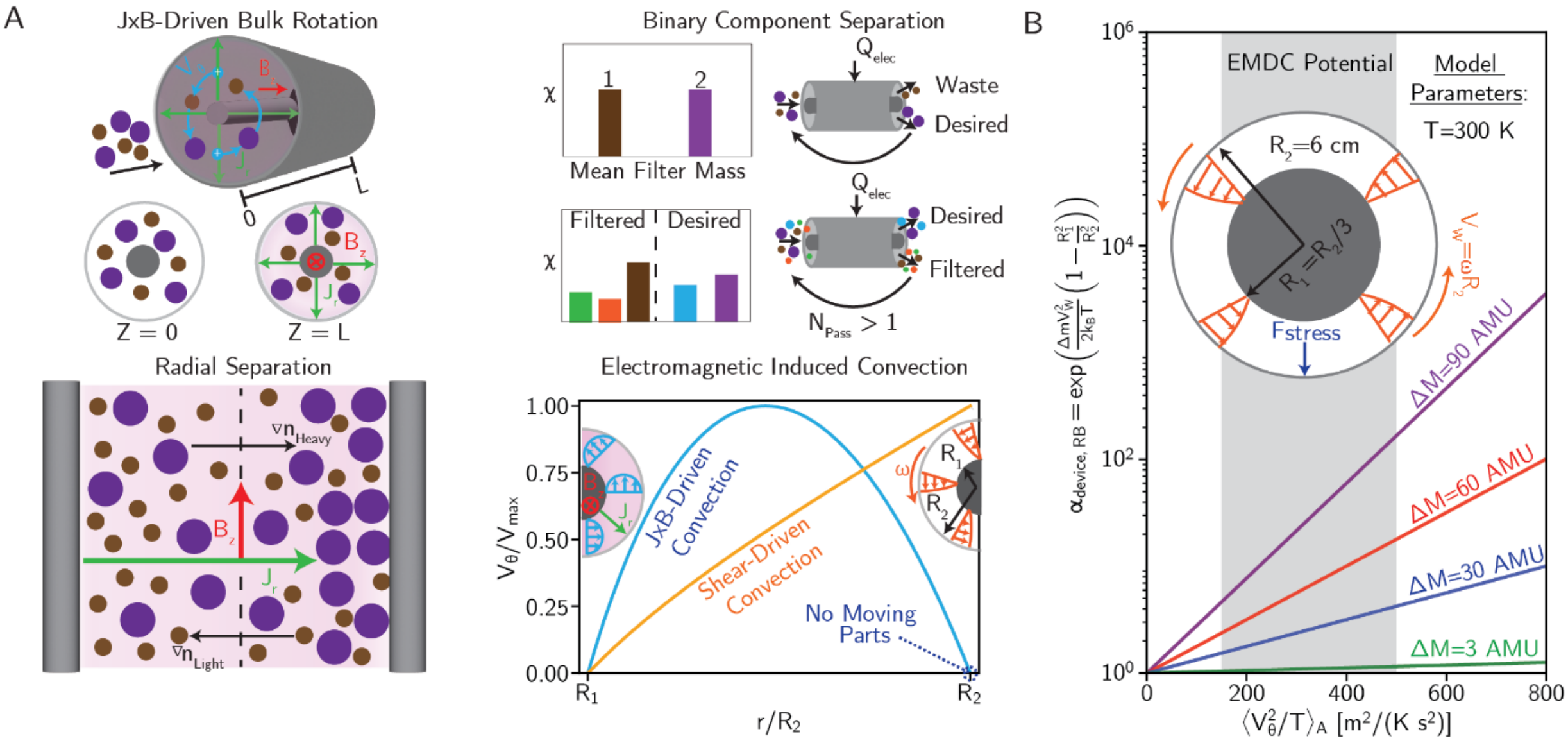

**Figure 1.** Rotational forcing enables mass separation. **(A)** A radial current density, $J_r$, crossed with an applied axial magnetic field, $B_z$, produces azimuthal Lorentz forcing and bulk rotation. The resulting centrifugal force and concentration gradients establish a mass distribution with heavier species pushed towards $R_2$ and lighter species towards $R_1$, enabling radial binary-component separation without a moving rotor. **(B)** Isothermal rigid-body annular device-level separation factor, $\alpha_{device,RB}$, plotted against the area-averaged parameter, $\langle V_\theta^2/T\rangle_A = \int_{R_1}^{R_2}(V_\theta^2/T)r\,dr \,/\, \int_{R_1}^{R_2} r\,dr$. The curves show that single-device radial separation increases strongly with both $\langle V_\theta^2/T\rangle_A$ and species mass difference, $\Delta m$.

The scaling of mass separation in SDCs is limited by the mechanism in which momentum is coupled into the flow. In these systems, a rotating wall produces large values of $\Delta\lambda$ near that boundary, but the rotational momentum must be transferred from the wall into the bulk fluid by viscous shear. As a result, $\Delta\lambda$ decays away from the rotating boundary. A volumetric forcing mechanism could instead sustain higher $\Delta\lambda$ values over a larger portion of the annulus and produce a greater integrated $\alpha_{device}$, even if the peak $\Delta\lambda$ value is lower than an SDC. This higher $\alpha_{device}$ would benefit low $\Delta m$ applications, such as $^{235}UF_6/^{238}UF_6$ enrichment and addressing tritium burn efficiency in fusion energy devices [7]. However, volumetrically distributed values of even moderate $\Delta\lambda$ extend the application space to high-throughput multi-component feedstock processing due to being able to sample a radial species distribution with finer mass resolution across the annulus. This is explained by the inner wall anchored separation factor, $\alpha_{R_1\to r} = \exp(\int_{R_1}^{r}(2\Delta\lambda/r)\,dr)$, where high $\alpha_{R_1\to r}$ values for an SDC exist for $r \sim R_2$, reducing the length scale over which the strongest mass transport occurs. This limits effective sampling to be near the outer wall rather than throughout the bulk. These considerations motivate alternative centrifuge concepts that generate volumetric forces which penetrate the bulk flow without rotor constraints.

Electromagnetically driven centrifuges (EMDCs) overcome the mechanical limits of shear-driven systems by replacing wall-driven rotation with volumetric Lorentz forcing $(\vec{J} \times \vec{B})$, where $\vec{J}$ is the current density, and $\vec{B}$ is the magnetic field. Azimuthal motion is generated by the interaction of current density and magnetic field with charged particles that produce bulk rotation throughout the fluid volume (Fig. 1A). Experiments have reported that these devices can achieve neutral azimuthal velocities of approximately 0.7-1.4 km/s in weakly ionized plasma centrifuges [8, 9], 7-13 km/s in pulsed plasma-chemical configurations [10], and flow speeds exceeding 1 km/s in fully ionized vacuum-arc centrifuges [11]. Beyond EMDCs, plasma-based mass separation has been explored using crossed-field mass filters [6], resonant ion-cyclotron isotope separation [12], RF-driven counter-current centrifuges [13], and transient laser-ablation effects [14]. These approaches demonstrate the flexibility of plasma separation physics, but many operate in low-pressure, highly ionized, resonant, or transient regimes that require substantial magnetic, RF, vacuum, or ion-collector infrastructure.

Weakly ionized EMDCs occupy a unique operating window, where low ionization fractions couple electromagnetic forcing into a dense neutral gas through collisions with the charged species that respond to the Lorentz force [8, 9, 15-17]. This process separates the plasma species by mass and utilizes higher pressures to facilitate ion-neutral collisions which drive momentum exchange between charged particles and neutral species. The higher densities due to the increased pressures prove favorable for high throughput applications. However, weakly ionized EMDC performance depends on sustaining sufficient ionization and thermalization to generate the current densities and collision frequencies that are required for large bulk velocities. When sustained, effective operation for mass separation requires mitigating the associated thermal dissipation mechanisms that work to undo the directed energy of the magnetized flow. The central question is therefore whether volumetric electromagnetic forcing in a weakly ionized EMDC can sustain sufficiently high $\lambda$ conditions across the annulus to yield competitive $\alpha_{R_1 \to r}$.

In this work, we show that weakly ionized EMDCs can access energetically favorable centrifugal regimes $(\lambda > 1)$ when electromagnetic forces, thermal dissipation, radial geometry, and magnetized transport are treated as coupled constraints. A two-temperature magnetohydrodynamic (2T-MHD) framework is developed and evaluated against literature and present facility measurements spanning abnormal-glow and arc regimes under low to high magnetic field conditions ($0.1 \leq \mathrm{B} \leq 0.57$ T), thereby establishing bounds on model validity. Following validation, the results demonstrate that the previously suggested $\lambda < 1$ viscous-dissipation limit [8] is not universal. The model predicts $\lambda > 1$ in select regimes ($7.5 \leq \mathrm{J} \leq 10$ kA/m$^2$, $0.4 \leq \mathrm{B} \leq 0.45$ T, $3 \leq R_2/R_1 \leq 4$). Joule heating and anisotropic transport set the dominant performance limits on $\alpha_{\mathrm{device}}$, while viscous dissipation is secondary. For the modeled $^{40}$Ar/$^{36}$Ar mixture, it is demonstrated that the highest $\alpha_{\mathrm{device}}$ values obtained (~1.33) do not correspond to the conditions that produce the largest $\lambda$ or area-averaged $\langle\lambda\rangle_{\mathrm{A}}$. Rather, it is the distribution of $\lambda/r$ that produces optimal $\alpha_{\mathrm{device}}$ conditions. This scaling is shown to favor EMDCs over SDCs, with the former able to approach or match the $\alpha_{\mathrm{device}}$ of the latter while exceeding SDC $\alpha_{R_1 \to r}$ over substantial portions of the annulus for lower $\langle\lambda\rangle_{\mathrm{A}}$. The source of the advantage is attributed to both the volumetric forcing and the ability to tune the MHD properties of the flow to reduce the influence of wall losses on the momentum balance. These results support further exploration of weakly ionized EMDC systems for high-throughput, multi-component feedstock processing applications.

## 2. 2T-MHD Framework

A 2T-MHD framework is used to model axisymmetric EMDCs under steady laminar flow conditions and validate their scaling of mass separation metrics. The model assumes the continuum approximation is valid throughout the annulus (Knudsen number, $\mathrm{Kn} \ll 0.1$) and time-varying magnetic fields remain negligible due to the rotating plasma ($\mathrm{R_m} \ll 1$). The applied current density is also assumed to induce fields much smaller than the applied field ($|B_{in}| \ll |B_z|$) by Ampere's law.

The plasma within EMDCs is composed of electrons, $\mathcal{N}_i$ ion species, and $\mathcal{N}_n$ neutral species, such that $\mathcal{S} = \mathcal{S}_e \cup \mathcal{S}_i \cup \mathcal{S}_n = \{e^-\} \cup \{s_{i1}, \ldots, s_{i\mathcal{N}_i}\} \cup \{s_{n1}, \ldots, s_{n\mathcal{N}_n}\}$. For the study of EMDCs, it is useful to distinguish heavy species, $\mathcal{S}_h = \mathcal{S}_i \cup \mathcal{S}_n$ and charged species, $\mathcal{S}_q = \mathcal{S}_e \cup \mathcal{S}_i$. The species lab-frame velocity, $\vec{V}_s = \vec{V} + \vec{w}_s$ is decomposed into the common mass-averaged bulk velocity, $\vec{V} = \left(\rho_h \vec{V}_h + \rho_e \vec{V}_e\right)/\rho$, and the species drift velocity, $\vec{w}_s$. Here, $\rho = \sum_{s\epsilon\mathcal{S}} \rho_s = \sum_{s\epsilon\mathcal{S}_h} \rho_s + \rho_e = \rho_h + \rho_e$ is the mixture mass density, and $\vec{V}_h = \sum_{s\epsilon\mathcal{S}_h} \rho_s \vec{V}_s/\rho_h$ is mass-averaged heavy species velocity. By definition $\langle \vec{w} \rangle_\rho = \sum_{s\epsilon\mathcal{S}} \rho_s \vec{w}_s/\rho = 0$, and the mean heavy particle mass is $\overline{m}_h = \rho_h/n_h = \sum_{s\epsilon\mathcal{S}_h} \chi_s^{(h)} m_s$, with the mole fraction, $\chi_s^{(h)} = n_s/n_h$ and number density, $n_h = \sum_{s\epsilon\mathcal{S}_h} n_s$.

The governing EMDC equations consist of global mixture mass and momentum conservation equations, with individual energy equations for the heavy species and electrons. The separate energy equations capture the nonequilibrium scaling of the heavy-species and electron temperatures, $T_h$ and $T_e$. These equations are coupled to obtain the mixture mass-averaged velocity, $\vec{V}$, heavy-species temperature, $T_h$, electron temperature, $T_e$, and pressure response that specify the centrifugal state of the EMDC [18-21]. The mathematical system is expressed as follows,

$$\nabla \cdot \left(\rho \vec{V}\right) = 0, \tag{1}$$

$$\rho \vec{V} \cdot \nabla \vec{V} = -\nabla \cdot \overleftrightarrow{P} + \rho_c \vec{E}^* + \vec{J} \times \vec{B}, \tag{2}$$

$$\rho_h \vec{V}_h \cdot \nabla e_{int,h} = -\nabla \cdot \vec{q}_h - \overleftrightarrow{P}_h : \nabla \vec{V}_h + \sum_{s\epsilon\mathcal{S}_i} \vec{J}_s \cdot \vec{E}^* + Q_{eh}^{el}, \tag{3}$$

$$\rho_e \vec{V}_e \cdot \nabla e_{int,e} = -\nabla \cdot \vec{q}_e - \overleftrightarrow{P}_e : \nabla \vec{V}_e + \vec{J}_e \cdot \vec{E}^* - Q_{eh}^{el} - \sum Q_p^{\text{Inelastic}}, \tag{4}$$

$$\vec{J} = \sum_{s\epsilon\mathcal{S}_q} \vec{J}_s = \sum_{s\epsilon\mathcal{S}_q} z_s e n_s \vec{w}_s = \sum_{s\epsilon\mathcal{S}_q} \begin{bmatrix} \sigma_{ps} & \sigma_{Hs} \\ -\sigma_{Hs} & \sigma_{ps} \end{bmatrix} \cdot \vec{\xi}_s, \tag{5}$$

$$\nabla \cdot \left(\rho_c \vec{V} + \vec{J}\right) = 0, \tag{6}$$

Here, the applied current density is assumed to flow radially between the inner electrode and outer electrode with radii, $R_1$ and $R_2$. Additionally, a no-slip boundary condition is assumed at the annulus walls, which enforces $\vec{V}(R_1) = \vec{V}(R_2) = 0$ and $V_r = 0$ per Eq. 1.

The second rank pressure tensor, $\overleftrightarrow{P}$, describes the sum of partial scalar pressure, $p_s$, and shear stress, $\overleftrightarrow{\tau}_s$ associated with each species $s\epsilon\mathcal{S}$, and is written as,

$$\overleftrightarrow{P} = \sum_{s\epsilon\mathcal{S}} \overleftrightarrow{P}_s = \sum_{s\epsilon\mathcal{S}} (p_s \mathbb{I} - \overleftrightarrow{\tau}_s) = \overleftrightarrow{P}_n + \overleftrightarrow{P}_i + \overleftrightarrow{P}_e = \overleftrightarrow{P}_h + \overleftrightarrow{P}_e. \tag{7}$$

The neutral and ion summed species pressure tensors are added to obtain the heavy species pressure, $\overleftrightarrow{P}_h = \overleftrightarrow{P}_n + \overleftrightarrow{P}_i$. This grouping assumes that the ion temperature, $T_i$ and neutral temperature, $T_n$ are treated to be in thermal equilibrium [18, 22], and represented by a single heavy temperature $T_h = T_n = T_i$. The total scalar pressure is the sum of the individual species pressures, $p = \sum_{s\epsilon\mathcal{S}} p_s = \sum_{s\epsilon\mathcal{S}} \rho_s k_B T_s/m_s$. Here $m_s$ is the particle mass of species s and $k_B$ is the Boltzmann constant. The electron and heavy-species temperatures, $T_e$ and $T_h$ are obtained through energy Eqs. 3 and 4, which use a monatomic ideal gas specific internal energy and constant volume specific heat closures, $e_{int,s} = 3T_s k_B/2m_s$ and $C_{Vs} = 3k_B/2m_s$.

To reduce the shear stress, $\overleftrightarrow{\tau} = \sum_{s\epsilon\mathcal{S}} \mu_s^{(4)} : \left(\nabla\vec{V}_s + \left(\nabla\vec{V}_s\right)^T\right)$ and heat flux $\vec{q} = -\sum_{s\epsilon\mathcal{S}} \overleftrightarrow{\kappa}_s \cdot \nabla T_s$ closures in the heavy-electron mixture, we recognize that each $s_q$ has its cross-field components of the fourth-rank viscosity tensor $\mu_{s_q}^{(4)}$, and the thermal conductivity tensor, $\overleftrightarrow{\kappa}_s$, suppressed following the Braginskii-type scaling of $\sim 1/(1 + \beta_{s_q}^2)$ [23-25]. Here, the magnetization is defined by the Hall parameter, $\beta_{s_q} = z_{s_q} e B_z / m_{s_q} \nu_m^{s_q}$ [24], where $z_{s_q}$ is the signed charge number, with elementary charge, e, and $\nu_m^{s_q}$ is the effective elastic momentum-relaxation frequency for $s_q$. The retained collision contributions depend on the species and the adopted collision closure [26]. Under the modeled discharge conditions, the Hall parameters satisfy $|\beta_e| \gg |\beta_{s_i}|$ for ion species $s_i \epsilon \mathcal{S}_i$, therefore the shear and heat flux terms can be reasonably approximated as $\sum_{s\epsilon\mathcal{S}} \overleftrightarrow{\tau}_s \sim \overleftrightarrow{\tau}_h$ and $\sum_{s\epsilon\mathcal{S}} \vec{q}_s \sim \vec{q}_h$.

The viscosities of the heavy species can be blended to form an effective heavy viscosity following the convention outlined in [18] and [26]. Under weakly ionized discharge conditions where $n_i/n_h \ll 1$, the heavy-species transport is neutral dominated. Therefore, the viscosity and viscous stress reduce to $\mu_h^{(4)} \sim \mu_n$ and $\overleftrightarrow{\tau}_h \sim \overleftrightarrow{\tau}_n$. Similarly, the thermal conductivity and heat flux reduce to $\overleftrightarrow{\kappa}_h \sim \kappa_n$ and $\vec{q}_h \sim \vec{q}_n$.

Following a Chapman-Enskog closure, the viscosity of a dilute monatomic neutral gas species, $s_n \epsilon \mathcal{S}_n$ is modelled as,

$$\mu_{s_n} = \frac{5\sqrt{\pi m_{s_n} k_B T_h}}{16\pi\sigma_{s_n}^2 \Omega^{(2,2)*}}, \tag{8}$$

where $\sigma_{s_n}$ is the Lennard-Jones collisional diameter for the neutral species. The collision integral, $\Omega^{(2,2)*}$, is a curve-fitted function [27, 28] of the reduced temperature, $T^* = k_B T_h/\varepsilon_{s_n}$, with $\varepsilon_{s_n}$ being the Lennard-Jones potential well depth of $s_n$ [26, 29]. The heavy thermal conductivity is related to the viscosity via [29],

$$\kappa_{s_n} = \mu_{s_n} \frac{15 k_B}{4 m_{s_n}}. \tag{9}$$

In the fluid frame the electric field is represented as $\vec{E}^* = \vec{E} + \vec{V} \times \vec{B}$, with $\vec{V} \times \vec{B}$ being the induced field from the conducting fluid moving normal to a magnetic field. We assume a singly ionized quasi-neutral plasma, and neglect sheaths in this model. Therefore, the net charge density, $\rho_c = \sum_{s\epsilon\mathcal{S}_q} z_{s_q} e n_{s_q}$ is zero and space-charge effects are omitted. Given this treatment, Eq. 6 reduces to $\vec{J}_r = \pm(J_{max} R_1/r)\hat{r}$ where the sign is determined by the assigned discharge polarity.

A locally imposed Saha relation is used as a bounded thermodynamic closure [18, 30] for the electron number density, $n_e$. This quantity is necessary to compute $\nu_m^{s_q}$, $\vec{F}_{s_q}$, and the elastic electron-heavy energy exchange, $Q_{eh}^{el} = 3 m_e n_e k_B \nu_m^e (T_e - T_h)/\overline{m}_h$ in Eqs. 3 and 4. Saha equilibrium is expressed as,

$$\frac{n_e n_a^{(c+1)}}{n_a^{(c)}} = \frac{2 Z_a^{(c+1)}}{Z_a^{(c)}} \left(\frac{2\pi m_e k_B T_e}{h^2}\right)^{3/2} \exp\left(-\frac{E_{ion,a}^{(c)}}{k_B T_e}\right), \tag{10}$$

where $n_a^{(c)}$ is the number density of the parent atomic or molecular species, a, in ionization state c, with $n_a^{(0)}$ denoting the neutral species. $E_{ion,a}^{(c)}$ is the energy required to ionize species a from state c to c+1, h is Planck's constant, and $Z_a^{(c)}$ is the internal partition function of species a in ionization state c [18]. In assuming a singly ionized plasma, we are considering $c = 0$ only. Nonequilibrium discharges can deviate from the Saha relation because of faster recombination mechanisms and using $T_e$ rather than $T_h$ in the absence of local thermal equilibrium (LTE) can overestimate ionization [30]. Eq. 10 is employed here to provide an approximate upper-limit estimate to capture centrifugal scaling. Similarly, in Eq. 4, inelastic excitation pathways, and detailed plasma-chemical rates, are also neglected [21, 30].

Fully ionized Spitzer-type transport closures have also been used in prior rotating plasma centrifuge analyses [8, 9], providing a high-ionization reference for comparison with the present Saha-derived conductivity. The relative scaling between Spitzer conductivity [31] and Saha-derived conductivity is demonstrated in Fig. 2 for a plasma produced by the electron-impact ionization reaction, $e^- + Ar \leftrightarrow Ar^+ + 2e^-$. As $T_e$ increases under the imposed Saha closure, electron-ion collisions increasingly determine the rate of electron momentum relaxation, and the conductivity approaches Coulomb-plasma Spitzer behavior. As discharge power density and collisional momentum exchange increase, the electron-heavy particle thermal equilibration becomes stronger, and LTE becomes a more reasonable approximation.

The conduction current is expressed with Ohm's law in Eq. 5 where the moving charged species interact with the applied magnetic field. This directional response results in an anisotropic conductivity tensor, $\overleftrightarrow{\sigma}$ [19, 32]. As stated, charged particles experience a magnetically induced reduction to their cross-field transport properties, and the magnitude of this reduction is captured by $\beta_{s_q}$. For an axially applied field, the radial (Pedersen), $\sigma_\mathrm{p}$, and azimuthal (Hall), $\sigma_\mathrm{H}$, conductivities are expressed as follows,

$$\sigma_\mathrm{p} = \sum_{s\epsilon\mathcal{S}_q} \sigma_{\mathrm{p}s} = \sum_{s\epsilon\mathcal{S}_q} \frac{\sigma_{\parallel s}}{1+\beta_s^2}, \tag{11}$$

$$\sigma_\mathrm{H} = \sum_{s\epsilon\mathcal{S}_q} \sigma_{\mathrm{H}s} = \sum_{s\epsilon\mathcal{S}_q} \frac{\sigma_{\parallel s}\beta_s}{1+\beta_s^2}. \tag{12}$$

The reduction to the cross-field drift motion due to $\beta_{s_q}$ is made explicit from both $\sigma_\mathrm{P}$ and $\sigma_\mathrm{H}$ being defined in terms of the axial (parallel) conductivity,

$$\sigma_\parallel = \sum_{s\epsilon\mathcal{S}_q} \sigma_{\parallel \mathrm{s}} = \sum_{s\epsilon\mathcal{S}_q} \frac{z_s^2 \mathrm{e}^2 \mathrm{n}_s}{\mathrm{m}_s \mathrm{v}_\mathrm{m}^s}. \tag{13}$$

In solving for these conductivity components, additional perturbative terms are found to influence the drift response and are represented in the generalized electric field for a charged species [19, 20, 32],

$$\vec{\xi}_{s_q} = \vec{\mathrm{E}}^* + \vec{\varepsilon}_{s_q} = \left(\vec{\mathrm{E}} + \vec{\mathrm{V}} \times \vec{\mathrm{B}}\right) - \left(\frac{1}{\mathrm{z}_{s_q}\mathrm{e}\mathrm{n}_{s_q}} \nabla \cdot \overleftrightarrow{\mathrm{P}}_{s_q} + \frac{\mathrm{m}_{s_q}}{\mathrm{z}_{s_q}\mathrm{e}} \vec{\mathrm{V}}_{s_q} \cdot \nabla \vec{\mathrm{V}}_{s_q}\right). \tag{14}$$

It follows that, in addition to the commonly shared $\vec{\mathrm{E}}^*$, each charged species experiences species-selective pressure and convective corrections, $\vec{\varepsilon}_{s_q}$. Under the assumed axisymmetric conditions, $E_\theta^* = 0$ [15]. The superscript star is reserved for the electric field transformed to the mass-averaged fluid frame, therefore both $\vec{\xi}_{s_q}$ and $\vec{\varepsilon}_{s_q}$ are left unstarred. In its unreduced form, Eq. 14 is formally implicit in $\vec{w}_{s_q}$ because $\vec{\mathrm{V}}_{s_q} = \vec{\mathrm{V}} + \vec{w}_{s_q}$, while Eq. 5 relates $\vec{w}_{s_q}$ to $\vec{J}_{s_q}$. The following ordering evaluates the drift-dependent contributions and reduces Eq. 14 to an explicit conductivity closure.

Within the regime of interest, the electron and ion pressure-gradient corrections are taken to be negligible. The charged-species viscous contribution, $\nabla \cdot \overleftrightarrow{\tau}_{s_q}$ is also neglected following the discussion preceding Eq. 8. To assess the inertial correction in Eq. 14 and close the lab-frame species velocity appearing in Eqs. 3 and 4, we consider the scaling of the drift velocity, $\vec{w}_s = w_{rs}\hat{r} + w_{\theta s}\hat{\theta}$. For all neutral species, $s_n$, the drift velocity is taken to be negligible, $\vec{w}_{s_n} = 0$. The charged species cannot be handled in the same way because Eq. 5 relates $\vec{w}_{s_q}$ to the current density, $\vec{J}_{s_q}$.

To approximate that the species are co-rotating, such that $V_{\theta s} \sim V_\theta$, we require $|V_\theta| \gg \left|\mathrm{w}_{\theta \mathrm{s_q}}\right|$. By summing the species azimuthal momentum equations over the charged species with electromagnetic forces balanced by elastic momentum exchange, we obtain the relation,

$$\frac{|\overline{\mathrm{w}}_{\theta q}|}{|V_\theta|} = \frac{\left|(\vec{\mathrm{J}} \times \vec{\mathrm{B}})_\theta\right|}{|V_\theta| \rho_q \nu_{m,eff}^{\rho_q}}, \tag{15}$$

where we have normalized by $|V_\theta|$. Here, we define the effective density-averaged momentum collision frequency for the charged species, $\nu_{m,eff}^{\rho_q} = \sum_{s\epsilon\mathcal{S}_q} \sum_{s_b \epsilon \mathcal{S}_n} \rho_s \nu_m^{ss_b} / \sum_{s\epsilon\mathcal{S}_q} \rho_s$ and the momentum transfer-weighted charged-neutral relative drift, $\overline{\mathrm{w}}_{\theta q} = \sum_{s\epsilon\mathcal{S}_q} \sum_{s_b \epsilon \mathcal{S}_n} \rho_s \nu_m^{ss_b} (\mathrm{w}_{\theta s} - \mathrm{w}_{\theta s_b}) / \sum_{s\epsilon\mathcal{S}_q} \sum_{s_b \epsilon \mathcal{S}_n} \rho_s \nu_m^{ss_b}$. In this work, we consider regimes where $|\overline{\mathrm{w}}_{\theta q}|/|V_\theta| \ll 1$, such that the momentum-transfer-weighted charged-neutral relative drift is small relative to the bulk azimuthal velocity. This motivates the co-rotation approximation of $V_{\theta s} \sim V_\theta$.

We cannot use a similar $|V_r| \gg \left|\mathrm{w}_{\mathrm{rs_q}}\right|$ relation for the radial drift speeds because Eq. 1 enforces $V_r = 0$ with the steady laminar axisymmetric flow, and Eqs. 5 and 6 demand a closure for $\mathrm{w}_{\mathrm{rs_q}}$. Expanding the convective term component-wise with these flow assumptions in Eq. 14 we obtain, $\left(\vec{\mathrm{V}}_{s_q} \cdot \nabla \vec{\mathrm{V}}_{s_q}\right)_r = w_{rs_q} dw_{rs_q}/dr - V_\theta^2/r$ and $\left(\vec{\mathrm{V}}_{s_q} \cdot \nabla \vec{\mathrm{V}}_{s_q}\right)_\theta = w_{rs_q}(dV_\theta/dr + V_\theta/r)$. We omit $\mathrm{m}_e \vec{\mathrm{V}}_e \cdot \nabla \vec{\mathrm{V}}_e / \mathrm{z}_e \mathrm{e}$ due to the small electron inertia. For the ions, the radial drift velocity is taken to be perturbative. Consequently, both $w_{rs_i} dw_{rs_i}/dr$ and $\left(\vec{\mathrm{V}}_{s_i} \cdot \nabla \vec{\mathrm{V}}_{s_i}\right)_\theta$ are neglected as non-leading-order contributions relative to electromagnetic and momentum exchange terms. This leaves $\vec{\varepsilon}_e \sim \vec{\varepsilon}_e^* = 0$, $\vec{\varepsilon}_{s_i} \sim \varepsilon_{rs_i}^* \hat{r} = \mathrm{m}_{s_i} V_\theta^2 / \mathrm{z}_{s_i} \mathrm{e} r \hat{r}$, and $\vec{\xi}_{s_q} \sim \xi_{rs_q}^* \hat{r}$. Here, we have added the star superscript to reflect the reduced fluid-frame form. With these reductions, we define $E_r^* = \left(J_r - \sum_{s\epsilon\mathcal{S}_i} \sigma_{\mathrm{p}s}\, \varepsilon_{rs_i}^*\right)/\sigma_{\mathrm{p}}$ and the charged-species radial and azimuthal current densities are evaluated as $\mathrm{J}_{\mathrm{rs_q}} = \sigma_{\mathrm{Ps_q}} \xi_{rs_q}^*$ and $\mathrm{J}_{\theta \mathrm{s_q}} = -\sigma_{\mathrm{Hs_q}} \xi_{rs_q}^*$. We can then use Eq. 5 to derive $w_{rs_q} = \mathrm{J}_{\mathrm{rs_q}} / \mathrm{z}_{s_q} \mathrm{e} n_{s_q}$. Using this scaling, both $\rho_{\mathrm{s_q}} \vec{\mathrm{V}}_{\mathrm{s_q}} \cdot \nabla \mathrm{e}_{int,\mathrm{s_q}}$ and $\overleftrightarrow{\mathrm{P}}_{\mathrm{s_q}} : \nabla \vec{\mathrm{V}}_{\mathrm{s_q}}$ terms are found to be non-leading order in Eqs. 3 and 4 and are neglected.

Determining mass separation in a multi-component fluid can be done by solving a separate set of $\mathcal{N}$ species momentum equations. Given $V_\theta$, $T_h$, $T_e$, and p from Eqs. 1-6, the partial pressure of each species $s\epsilon\mathcal{S}$ that co-rotates at $V_\theta$ can be found using the species momentum equation,

$$\rho_s \vec{\mathrm{V}}_s \cdot \nabla \vec{\mathrm{V}}_s = \rho_{c,s} \vec{\mathrm{E}}^* + \vec{J}_s \times \vec{\mathrm{B}} - \nabla \cdot \overleftrightarrow{\mathrm{P}}_s - \vec{F}_s. \tag{16}$$

Here, $\rho_{c,s} = \mathrm{z}_s e \mathrm{n}_s$ and $\vec{J}_s$ are the charge and current density of the species. Following the simplifications made, the species lab frame velocity is expressed as, $\vec{V}_s = w_{rs}\hat{r} + V_\theta \hat{\theta}$. The collisional momentum exchange term, $\vec{F}_s = \sum_{s'\epsilon\mathcal{S}} \vec{F}_{ss'} = \sum_{s'\epsilon\mathcal{S}} m_s n_s \nu_m^{ss'} (\vec{V}_s - \vec{V}_{s'})$ on the right-hand side of Eq. 16 is not present in Eq. 2 when summing the species equations. This is because the collisional terms cancel via $\vec{F}_{ss'} = -\vec{F}_{s's}$ to satisfy global momentum conservation. A complete description of the reduced equations is given in Appendix A.

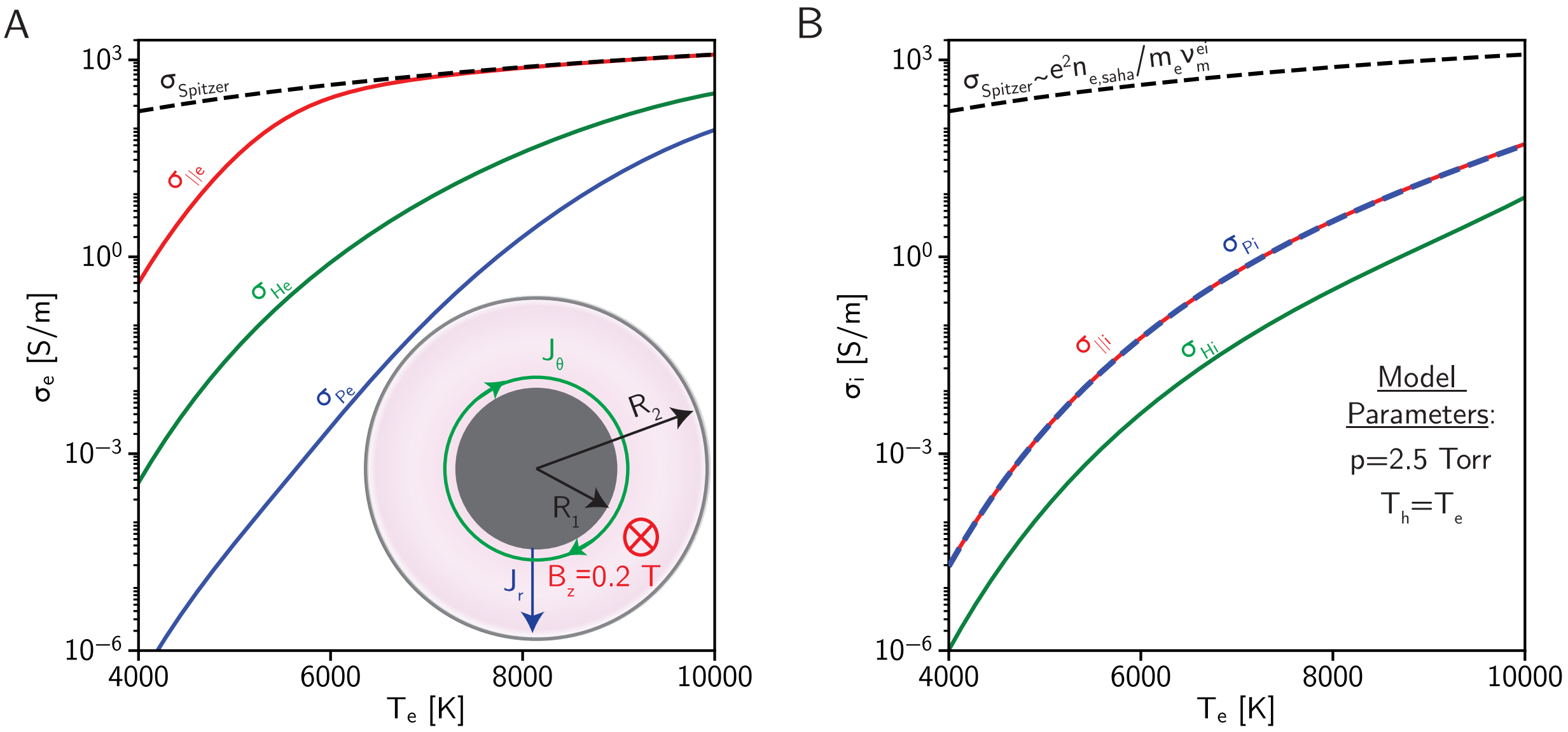

**Figure 2.** Temperature-dependent anisotropic conductivity components used in the 2T-MHD model under LTE conditions with charged species, $\mathcal{S}_q = \{e^-, Ar^+\}$. **(A)** Electron conductivity components as functions of $T_e$, including the parallel conductivity, $\sigma_{\parallel e}$, Pedersen conductivity, $\sigma_{Pe}$, Hall conductivity, $\sigma_{He}$, and the corresponding $\sigma_{Spitzer}$ reference. The inset cartoon illustrates the imposed radial current density, $J_r$, axial magnetic field, $B_z$, and induced azimuthal Hall current $J_\theta$ in the coaxial EMDC geometry. **(B)** Ion conductivity components evaluated over the same $T_e$ domain. The comparison shows how magnetization separates the parallel, Pedersen, and Hall responses for each charged species. Due to $|\beta_e| \gg |\beta_{Ar^+}|$, $\sigma_P$ is ion dominated over the majority of the $T_e$ values, while $\sigma_H$ is dictated primarily by the electrons.

### 2.1. EMDC Scaling Metrics

Nondimensional scaling metrics arise from the 2T-MHD model that indicate the performance of EMDCs. Critical to the analysis in this work, the radial total pressure gradient in Eq. 2 scales with the ratio of directed bulk rotational energy to thermal energy by substituting,

$$\lambda_h = \frac{\overline{m}_h V_\theta^2}{2k_B T_h}, \tag{17}$$

into the centrifugal term obtained from expanding, $\left(\rho \vec{V}_\theta \cdot \nabla \vec{V}_\theta\right)_r$. The 2T-MHD model captures the scaling, $\lambda_h$, which indicates the local centrifugal state of the bulk fluid. Additionally, an analogous quantity, $\lambda_s = m_s V_\theta^2 / 2k_B T_s$ is the species-specific version of Eq. 17 which emerges from Eq. 16.

A second scaling metric, the Hartmann number, Ha, measures the relative strength of electromagnetic momentum coupling and viscous momentum diffusion across a characteristic length scale. It emerges from the azimuthal momentum balance between the Lorentz-force response associated with the cross-field motion of a conducting magnetized fluid and viscous diffusion as,

$$\mathrm{Ha} = B_z \Delta R \sqrt{\frac{\sigma_P}{\mu_h}}. \tag{18}$$

In this work, Ha serves as a measure of the uniformity of the annular $V_\theta$ profile. When Ha is small, viscous diffusion and wall losses dominate the redistribution of angular momentum. Under these conditions, the radial distribution of $V_\theta$ is sharply tapered at the boundaries and a broad, nearly uniform core profile does not develop. When Ha is large, electromagnetic forces control the bulk momentum balance, allowing the imposed $\left(\vec{J} \times \vec{B}\right)_\theta$ to sustain a more radially uniform $V_\theta$ distribution across the annulus [33, 34].

### 2.2. Two-Component Mass Separation

The 2T-MHD model can be simplified to determine separation scaling for two parent heavy species, a, rotating at the mass-averaged speed, $V_\theta$ and characterized by the heavy temperature, $T_h$. Summing the radial form of Eq. 16 for the ion and neutral forms of a desired parent species, and subtracting the sum of another parent species produces,

$$\frac{d}{dr}\ln\frac{p_1}{p_2} = \frac{2\Delta\lambda_{1,2}}{r} + \frac{1}{p_1}\Big((\nabla\cdot\overleftrightarrow{\tau}_1)_r + \rho_{c,s_{i1}}E_r^* + \big(\vec{J}_{s_{i1}}\times\vec{B}\big)_r - \big(\vec{F}_1\big)_r\Big) \\ - \frac{1}{p_2}\Big((\nabla\cdot\overleftrightarrow{\tau}_2)_r + \rho_{c,s_{i2}}E_r^* + \big(\vec{J}_{s_{i2}}\times\vec{B}\big)_r - \big(\vec{F}_2\big)_r\Big), \tag{19}$$

where $\Delta\lambda_{1,2} = \Delta m_{1,2}V_\theta^2/(2k_BT_h)$ and $\Delta m_{1,2} = m_1 - m_2$. Here, we have introduced the parent-species-specific sums, $m_a = \sum_{s\epsilon\mathcal{S}_a}\rho_s/\sum_{s\epsilon\mathcal{S}_a}n_s$, $p_a = \sum_{s\epsilon\mathcal{S}_a}p_s$, $\overleftrightarrow{\tau}_a = \sum_{s\epsilon\mathcal{S}_a}\overleftrightarrow{\tau}_s$, and $\vec{F}_a = \sum_{s\epsilon\mathcal{S}_a}\vec{F}_s$, where $\mathcal{S}_a$ includes the ion and neutral forms of parent species, a. $\Delta\lambda_{1,2}$ connects the velocity and temperature fields from the 2T-MHD model to mass separation for two parent species.

For the case of an isotopic $^{40}Ar/^{36}Ar$ mixture, where $\mathcal{S}_1 = \{^{40}Ar, ^{40}Ar^+\}$ and $\mathcal{S}_2 = \{^{36}Ar, ^{36}Ar^+\}$, only the centrifugal $\Delta\lambda_{1,2}$ term is found to be significant for the weakly ionized conditions considered in this study. Integrating Eq. 19 between two radial points, $r_1$ and $r_2$ gives the radial separation factor,

$$\alpha_{r_1\to r_2} = \frac{\left(\frac{\chi_1}{\chi_2}\right)_{r_2}}{\left(\frac{\chi_1}{\chi_2}\right)_{r_1}} \sim \exp\left(2\Delta m_{1,2}\int_{r_1}^{r_2}\left(\frac{\lambda_h}{\bar{m}_h r}\right)dr\right). \tag{20}$$

Here, we have used the relation, $\Delta\lambda_{1,2} = \Delta m_{1,2}\lambda_h/\bar{m}_h$, and have written the partial pressures in terms of mole fraction as $p_1 = \chi_1 p_h$ and $p_2 = \chi_2 p_h = (1-\chi_1)p_h$, where $\chi_a = \sum_{s\epsilon\mathcal{S}_a}n_s/n_h$. For generality, $\bar{m}_h$ is retained inside the integrand. For the dilute isotopic mixture considered here, $\bar{m}_h \sim m_{Ar}$ throughout the annulus so the spatial variation of $\lambda_h/r$ dictates $\alpha_{r_1\to r_2}$. When referenced from the inner wall, Eq. 20 becomes $\alpha_{R_1\to r}$, and when integrated across the annulus yields the device separation factor, $\alpha_{device} = \alpha_{R_1\to R_2}$.

Eq. 20 provides the argument for why radial separation is not controlled by the peak value of $\Delta\lambda_{1,2}$ or the region near the outer wall as is optimized by SDCs. Instead, enrichment depends on how $\lambda_h/r$ is distributed radially between two points, $r_1$ and $r_2$. This fundamental concept is the source of the advantage a volumetrically driven flow has for mass separation by sustaining a higher $\lambda_h/r$ throughout a greater portion of the annular gap. If $\vec{J}\times\vec{B}$ can be optimized to produce an energetically favorable plasma with $\langle\lambda_h\rangle_A > 1$, while maintaining a state that minimizes momentum losses to the walls through $\langle Ha\rangle_A > 1$, EMDCs can surpass SDC values of $\alpha_{R_1\to r}$ over substantial portions of the annulus and, for favorable operating conditions and geometry, approach or match the corresponding $\alpha_{device}$.

To isolate the factors dictating Eq. 20 in an EMDC, radial pressure and mole-fraction profiles are shown in Fig. 3. Using a $\mathcal{S} = \{Ar, Ar^+, e^-\}$ composition, Eqs. 1-6 are solved to obtain $V_\theta$, $T_h$, and $p_h$ with $\bar{m}_h \sim m_{Ar}$. Then, for the $^{40}Ar/^{36}Ar$ mixture with a light-species feed mole fraction of $\chi_{L,F}$ ~ 0.33%, these calculated quantities are used in post-processing to determine the partial pressures and composition profiles across the annulus (Eq. 19). These profiles are constrained by axial mass conservation, with a prescribed flow rate and composition specifying the inlet feed conditions, and the outer wall is set at the total feed pressure, $p_f$ = 2.5 Torr. The EMDC case considered featured $R_2 = 6$ cm, $R_2/R_1 = 3$, $J_{max} = 2$ kA/m$^2$, and $B_z = 0.2$ T. Heavy-particle temperature is assumed to be $T_h(R_1) = T_h(R_2) = 300$ K, representing actively cooled coaxial electrodes, which provides a practical ideal thermal limit for studying the scaling of $\lambda_h$ in EMDCs.

Three electromagnetic configurations were compared: positive polarity ($\hat{J}_r = \hat{r}$) without the radial Lorentz force $\big(\vec{J}\times\vec{B}\big)_r$, negative polarity ($\hat{J}_r = -\hat{r}$) with $\big(\vec{J}\times\vec{B}\big)_r$, and $\hat{J}_r = \hat{r}$ with $\big(\vec{J}\times\vec{B}\big)_r$. These isolated conditions allow us to evaluate how they each influence annular pressure distributions and multi-component composition profiles within an EMDC.

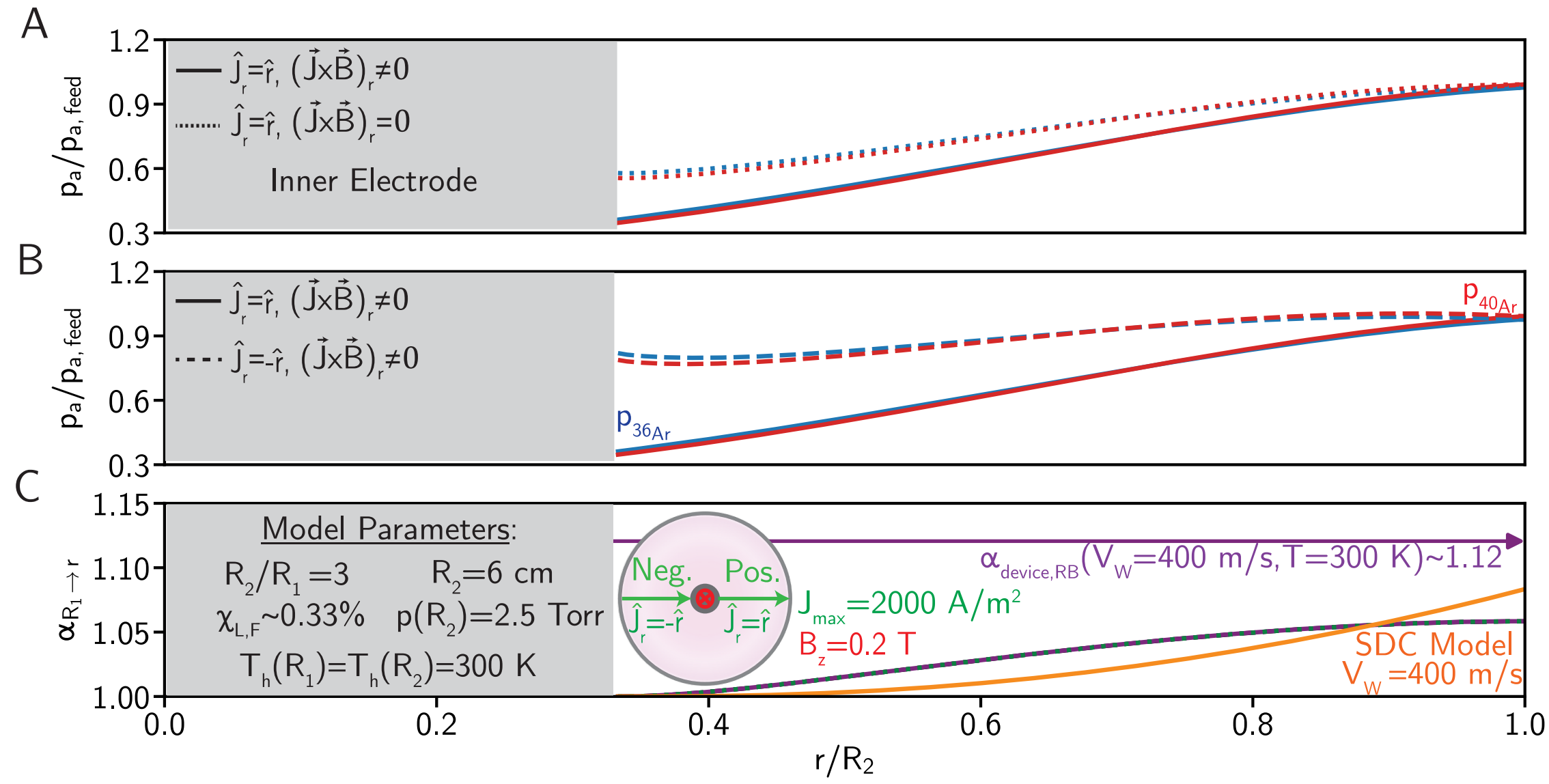

**Figure 3.** Two-component radial enrichment for a $^{40}Ar/^{36}Ar$ mixture **(A)** Radial partial-pressure profiles normalized by the feed pressure for $^{40}Ar$ and $^{36}Ar$ under $\hat{J}_r = \hat{r}$ discharge conditions with $(\vec{J} \times \vec{B})_r$ and $(\vec{J} \times \vec{B})_r = 0$. The $(\vec{J} \times \vec{B})_r$ contribution modifies the total pressure evolution, with the direction of the pressure shift being opposite the polarity direction. **(B)** Radial partial-pressure profiles of $\hat{J}_r = -\hat{r}$, $\hat{J}_r = \hat{r}$, and $(\vec{J} \times \vec{B})_r \neq 0$. The polarity of the discharge modifies the direction of the effect by $(\vec{J} \times \vec{B})_r$, and the curvature of the pressure profile. **(C)** Corresponding separation factor, $\alpha_{R_1 \to r}$, radial profiles. The composition ratio is set by the integrated $\Delta\lambda_{1,2}/r$ profile, therefore $(\vec{J} \times \vec{B})_r$ and polarity effects produce small variations in $\Delta\lambda_{1,2}$. The SDC model is included with a wall speed of 400 m/s to compare against the EMDC result, and the $\alpha_{device,RB}$ value is shown to illustrate how the SDC model deviates from the isothermal rigid body calculation traditionally used in centrifuge literature.

The radial Lorentz force modified the radial partial-pressure profiles within an EMDC (Fig. 3A). Relative to the case with $(\vec{J} \times \vec{B})_r = 0$, the profiles with $(\vec{J} \times \vec{B})_r \neq 0$ shifted the partial pressures of both species in the direction opposite the polarity direction. In Fig. 3B, the negative polarity resulted in the partial pressure curves continuing the trend of shifting in the direction opposite the applied polarity, and the curvature was altered relative to the positive polarity. Despite these partial pressure differences due to polarity and $(\vec{J} \times \vec{B})_r$, the minor variation among the EMDC curves in Fig. 3C indicates that these terms have a negligible effect on the $V_\theta$ and $T_h$ profiles, and therefore the effect on $\Delta\lambda_{1,2}$ can be neglected. That said, in regimes where the degree of ionization is significant, and the ion-specific $(\vec{J}_{s_i} \times \vec{B})_r$ terms in Eq. 19 cannot be neglected, it is possible that a difference in response to the radial Lorentz force between the species could influence separation scaling. Those regimes are beyond the scope of this study for weakly ionized EMDCs.

An SDC model (Appendix B) with wall speed $V_w = 400$ m/s is overlaid in Fig. 3C to compare the radial separation produced by both forcing mechanisms. The SDC $\alpha_{R_1 \to r}$ profile decays away from the rotating outer wall but ultimately yields an $\alpha_{device}$ approximately 0.02 greater than those of the EMDC cases. However, within the $0.4 \leq r/R_2 \leq 0.8$ annulus region, the EMDC produces greater $\alpha_{R_1 \to r}$ values than the SDC, with the difference near $r/R_2 \sim 0.6$ also yielding $|\Delta\alpha_{R_1 \to r}| \sim 0.02$. The $\alpha_{device,RB}(V_w = 400$ m/s, $T = 300$ K) value is included to illustrate that this result is not due to the SDC model uncharacteristically scaling the magnitude of $\alpha_{device}$. It demonstrates that the canonical isothermal rigid body device separation factor metric deviates modestly ($|\Delta\alpha_{device}| \sim 0.04$) from the SDC model prediction. The higher $\alpha_{R_1 \to r}$ values of the EMDC relative to the SDC across most of the annulus depict the advantage of volumetric forcing for mass separation explicitly. The remainder of the paper seeks to validate the 2T-MHD model and use it to identify pathways to further enhance $\Delta\alpha_{R_1 \to r}$ in favor of the EMDC and produce $\alpha_{device}$ values matching those achieved by SDCs operating near maximum $V_w$.

### 2.3. Viscous Dissipation Scaling in EMDCs

Simulations were performed using the 2T-MHD model to quantify how Joule heating $(\vec{J} \cdot \vec{E}^*)$, viscous dissipation $(\overleftrightarrow{\tau} : \nabla\vec{V})$, radial Lorentz forcing $(\vec{J} \times \vec{B})_r$, and cross-field transport anisotropy ($\beta_{s_q}$) affect EMDC performance. To isolate the effects of different physics, $\vec{J}\cdot\vec{E}^*$, $\overleftrightarrow{\tau}:\nabla\vec{V}$, $(\vec{J}\times\vec{B})_r$, and $\beta_{s_q}$ terms were toggled. The representative cases in Fig. 4 use $\mathcal{S} = \{Ar, Ar^+, e^-\}$ with p($R_2$) = 2.5 Torr, $R_2$ = 6 cm, $R_2/R_1$=3, $B_z$ = 0.2 T, and $J_{max}$ = 4000 A/m$^2$. We are not using the two-component feed like in Fig. 3 because we are ultimately concerned with the scaling of $\lambda_h$, and given a dilute mixture with argon as the majority species, the trace component is assumed to have negligible effect on $V_\theta$ and $T_h$ scaling.

When all terms are included ($\overleftrightarrow{\tau}:\nabla\vec{V} \neq 0, (\vec{J}\times\vec{B})_r \neq 0, \beta_{s_q} \neq 0, \vec{J}\cdot\vec{E}^* \neq 0$) we obtain the baseline EMDC condition. In this case, $T_e$ ~ 8000 K with a comparatively uniform profile because it is determined locally by the balance between $\vec{J}_e\cdot\vec{E}^*$ and $Q_{eh}^{el}$ with $\nabla\cdot\vec{q}_e$ neglected. The $T_h$ profile peaks at ~3300 K with a parabolic shape bounded by the imposed $T_h$ = 300 K wall conditions. Similarly, $V_\theta$ satisfies the no-slip boundary requirement with maximum speeds of ~750 m/s. These combined $V_\theta$ and $T_h$ profiles produce local $\lambda_h$ values up to 0.6.

The contribution of each physical mechanism to this response was determined by repeating the simulations with individual terms removed. When $\vec{J}\cdot\vec{E}^* = 0$, the model predicts substantially larger rotation of $V_\theta$ ~ 1.3-1.4 km/s, while $T_h$ approaches ~ 1000 K. This produces a broad region where $\lambda_h > 1$, with peaks near $\lambda_h$~2.4. Therefore, contrary to the claims of prior EMDC literature [8,9], $\overleftrightarrow{\tau}:\nabla\vec{V}$ alone does not enforce $\lambda_h < 1$ under these conditions. Instead, the main thermal penalty comes from $\vec{J}\cdot\vec{E}^*$ coupled with magnetized cross-field transport. This claim is consistent with the $\beta_{s_q} = 0, (\vec{J}\times\vec{B})_r = 0, \vec{J}\cdot\vec{E}^* \neq 0$ case, where removing the magnetized suppression of $\sigma_P$ lowers $\vec{J}\cdot\vec{E}^*$, which reduces $T_e$ to approximately 4000 K and leaves $T_h$ close to the $\vec{J}\cdot\vec{E}^* = 0$ case. Since $\vec{J}\cdot\vec{E}^* \sim J_r^2/\sigma_P$, magnetization increases the cost of driving radial current by reducing $\sigma_P$ relative to $\sigma_\parallel$.

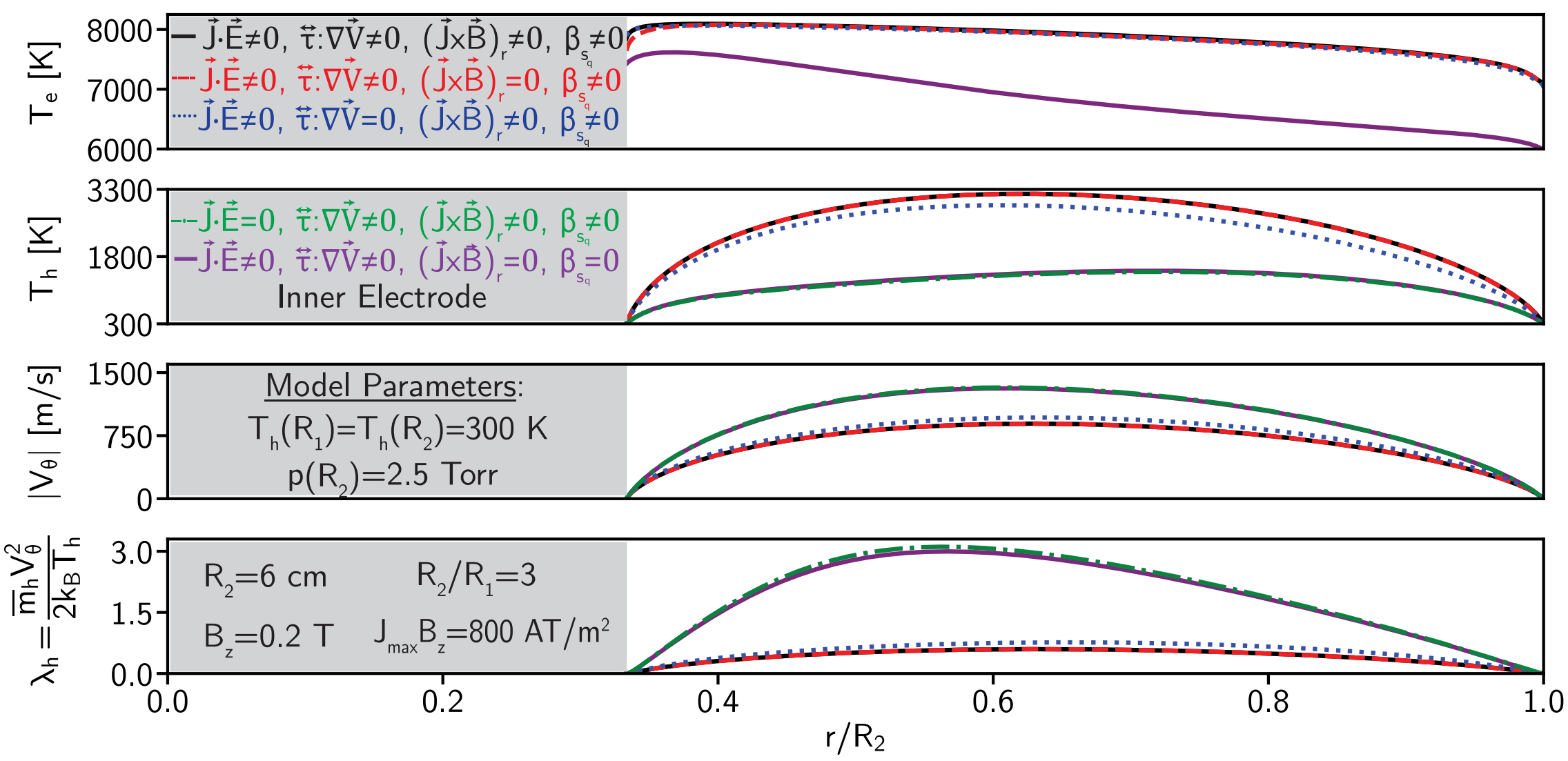

**Figure 4.** EMDC profiles used to isolate the roles of $\vec{J}\cdot\vec{E}^*$, $\overleftrightarrow{\tau}:\nabla\vec{V}$, $(\vec{J}\times\vec{B})_r$, and $\beta_{s_q}$ in setting $\lambda_h$. From top to bottom, the panels show the electron temperature, $T_e$, heavy-particle temperature, $T_h$, azimuthal velocity magnitude, $|V_\theta|$, and $\lambda_h$, as functions of radius. The cases are referenced to the full anisotropic model ($\overleftrightarrow{\tau}:\nabla\vec{V} \neq 0, (\vec{J}\times\vec{B})_r \neq 0, \beta_{s_q} \neq 0, \vec{J}\cdot\vec{E}^* \neq 0$) by toggling off the different physics terms. This exercise reveals that the dominant thermal penalty is $\vec{J}\cdot\vec{E}^*$ amplified by the magnetization-induced suppression of $\sigma_P$, rather than $\overleftrightarrow{\tau}:\nabla\vec{V}$.

### 2.4. Geometric and Thermal Scaling in EMDC Devices

Radial geometry controls EMDC performance by increasing radial integration length, $\Delta R = R_2 - R_1$ and promoting bulk flow conditions where the momentum redistribution is not dominated by viscous wall losses. The latter is correlated with high values of $\langle Ha\rangle_A$ and states with higher $\langle Ha\rangle_A$ generally exhibit higher $\langle\lambda_h\rangle_A$ through increased rotational speeds and decreased thermal dissipation. However, since $\alpha_{device}$ depends on the radial integral of $\lambda_h/r$, the best geometry is not the one with the largest local or area-averaged $\lambda_h$. The useful design target is instead determined by a combination of $\Delta R$, $\langle Ha \rangle_A$, and $\langle \lambda_h \rangle_A$.

Simulations were performed under fixed electromagnetic force ($B_z = 0.25$ T and $J_{max}B = 1250$ AT/m$^2$) conditions to study the correlation between geometry, $\langle Ha \rangle_A$, and $\langle \lambda_h \rangle_A$ (Fig. 5). In Fig. 5A, $R_2$ is swept from 1 to 6 cm and $R_2/R_1$ from 1.5 to 10. The results show that $\langle Ha \rangle_A$ increases strongly with $R_2$, while the aspect ratio has an intermediate optimum near $2 \le R_2/R_1 \le 6$. This behavior follows directly from Eq. 18, where larger devices increase $\Delta R$, but large aspect ratios reduce the ability of the imposed force to produce sufficient conductivity in the plasma.

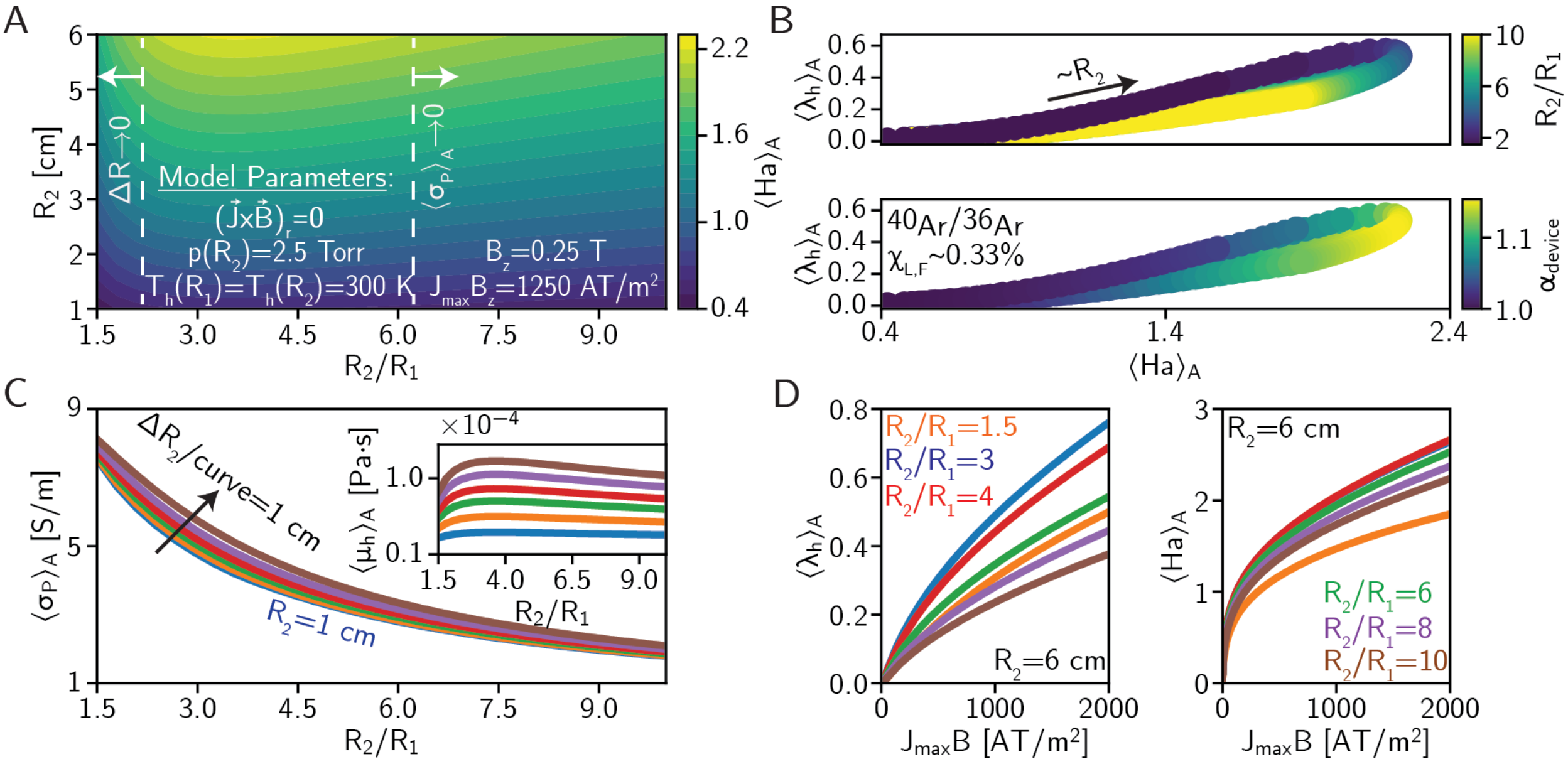

**Figure 5.** Scaling of $\langle Ha \rangle_A$, $\langle \lambda_h \rangle_A$, and $\alpha_{device}$ due to radial geometry and electromagnetic forces. **(A)** Area-averaged Hartmann number, $\langle Ha \rangle_A$, as a function of outer radius, $R_2$, and aspect ratio, $R_2/R_1$. This shows that Ha increases with reactor size but has an intermediate aspect ratio optimum. **(B)** Relationship between $\langle Ha \rangle_A$, $\langle \lambda_h \rangle_A$, $R_2/R_1$, and $\alpha_{device}$. The maximum $\alpha_{device}$ does not occur solely at the maximum $\langle \lambda_h \rangle_A$ condition because separation depends on the radially integrated $\lambda_h/r$ distribution, which is correlated with $\langle Ha \rangle_A$. **(C)** $\langle \sigma_P \rangle_A$ and $\langle \mu_h \rangle_A$ scaling with aspect ratio, showing that the low $R_2/R_1$ regime is limited by $\Delta R$ while the high $R_2/R_1$ regime is limited by reduced $\langle \sigma_P \rangle_A$. **(D)** Dependence of $\langle \lambda_h \rangle_A$and $\langle Ha \rangle_A$ on $J_{max}B$ for select $R_2/R_1$. These plots demonstrate that the electromagnetic force scaling hierarchy of $\langle \lambda_h \rangle_A$ based on geometry is not mirrored in the $\langle Ha \rangle_A$ scaling.

The geometric dependency of Ha is shown explicitly in Fig. 5C by showing how $\langle \sigma_P \rangle_A$ and $\langle \mu_h \rangle_A$ vary with aspect ratio for different $R_2$. Increasing $R_2$ increases $\Delta R$ and generally improves $\langle Ha \rangle_A$. At low aspect ratio, $R_2/R_1 < 2$, the electromagnetic force is focused on a smaller $\Delta R$, therefore the transport ratio $\langle \sqrt{\sigma_P/\mu_h} \rangle_A$ is favorable, but the annular gap is too small for the flow to develop away from viscous wall losses. At high aspect ratio, $R_2/R_1 > 6$, the larger $\Delta R$ cannot compensate for the reduction in $\langle \sigma_P \rangle_A$ due to the electromagnetic force being less focused in the gap. The intermediate range, $2 \le R_2/R_1 \le 6$ therefore, gives the best compromise between $\Delta R$ and sustaining a transport state favorable for promoting volumetrically redistributed flow.

Simulations were then performed to connect the scaling of $\langle \lambda_h \rangle_A$ with the $\langle Ha \rangle_A$ and aspect ratio scaling shown in Fig. 5A (Fig. 5B). For a $^{40}Ar/^{36}Ar$ mixture with $\chi_{L,F} \sim 0.33\%$, $\alpha_{device}$ is maximized when $\langle Ha \rangle_A$ and $\langle \lambda_h \rangle_A$ are optimized together (Fig. 5B). Fig. 5D demonstrates that the preferred aspect ratio depends on the applied electromagnetic force. At low $J_{max}B$, the curves are similar, but as the magnitude of the force increases, the $R_2/R_1 = 3$ and 4 cases begin to outperform the smaller and larger aspect ratios.

The ability of EMDCs to leverage $\langle Ha \rangle_A$ scaling for improved $\langle \lambda_h \rangle_A$ and $\alpha_{device}$ is unique to their use of the $\vec{J} \times \vec{B}$ force to drive rotation. This contrasts with SDCs, which lack the ability to tune the flow properties with boundary conditions and electromagnetic forcing to promote a higher integrated value of $\lambda_h/r$ through Ha scaling. However, SDCs do enjoy the initial advantage of their thermal dissipation being restricted to $\overleftrightarrow{\tau}:\nabla\vec{V}$ and generating high $\lambda_h$ values near their rotating outer wall. This distinction is made explicit for $R_2 = 6$ cm and $R_2/R_1 = 2$ to 4 in Fig. 6. The EMDC cases use $J_{max} = 2\ \mathrm{kA/m^2}$ and $B_z = 0.2$ T, while the SDC reference uses an outer-wall speed equal to the argon isentropic sound speed that is evaluated at $T_{h,min} = 300$ K. The EMDC develops larger $T_h$ because the $\vec{J} \times \vec{B}$ necessary for operation produces $\vec{J} \cdot \vec{E}^*$, in addition to the $\overleftrightarrow{\tau}: \nabla\vec{V}$ that is the only thermal dissipation mode in an SDC (Fig. 6A). Despite this heating penalty, the EMDC produces larger $V_\theta$ over more of the annulus (Fig. 6B) because the forcing is volumetric and the stationary-wall boundary conditions produce a parabolic profile. As $R_2/R_1$ grows from 2 to 4, $V_\theta$ increases because the wall losses are less influential on the bulk flow set by the electromagnetic force. The corresponding distribution of $\lambda_h$ is shown in Fig. 6C, and the EMDC curves show a clear trend of covering more of the annular gap with higher $\lambda_h$ values than the SDC curves as $R_2/R_1$ increases in the range considered. Therefore, to maximize $\Delta\alpha_{R_1 \to r}$ in favor of the EMDC and produce $\alpha_{device}$ values matching high-performance SDCs, the $R_2/R_1$ must be tuned to best sustain the integrated $\lambda_h/r$ profile. This choice is dependent on $\Delta R$ and the $\langle Ha \rangle_A$, and $\langle \lambda_h \rangle_A$ scaling discussed in Fig. 5.

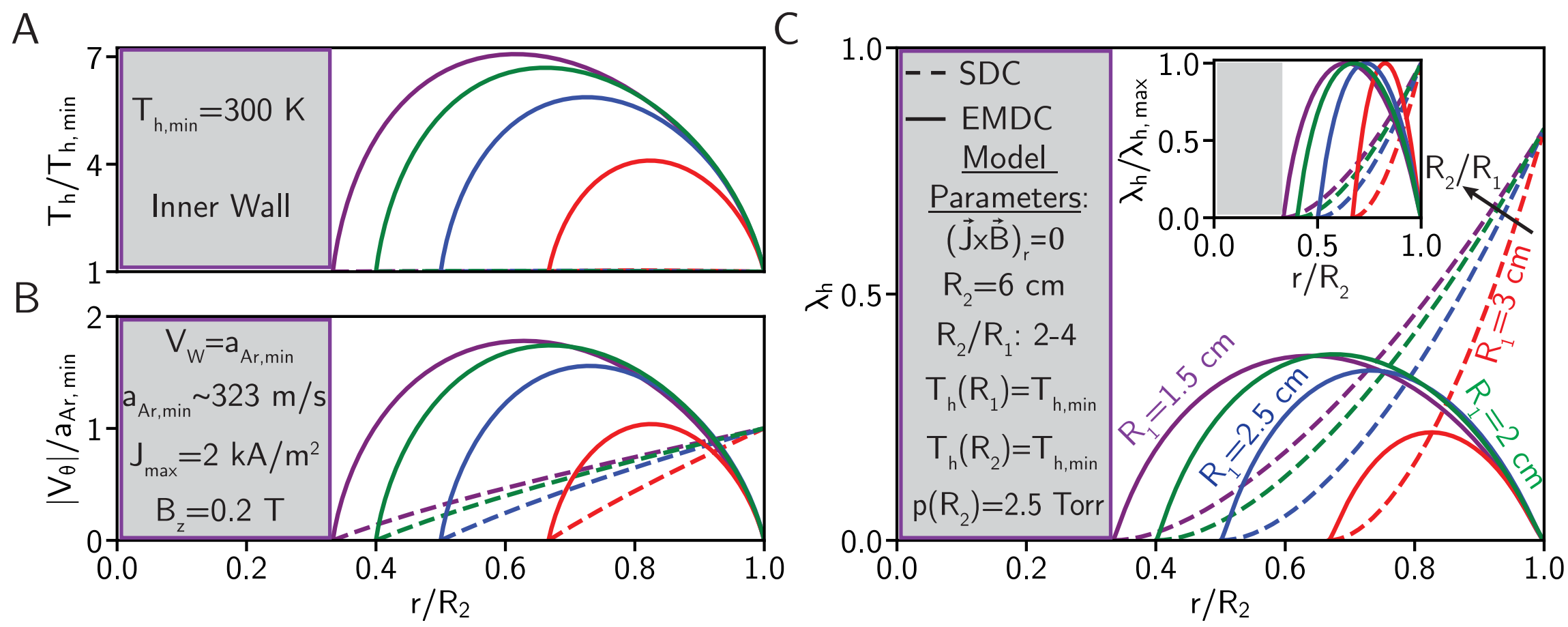

**Figure 6.** Comparison of EMDC and SDC radial profiles for $R_2 = 6$ cm and $R_2/R_1 = 2$ to 4. **(A)** Heavy-particle temperature normalized by $T_{h,min} = 300$ K. **(B)** Azimuthal velocity normalized by the argon isentropic sound speed at $T_{h,min}$. **(C)** Corresponding $\lambda_h$ profiles, with the inset showing each curve normalized by its own maximum value. The SDC produces the largest peak $\lambda_h$ near the rotating wall, but the EMDC sustains a broader $\lambda_h$ distribution because momentum is introduced volumetrically through $\vec{J} \times \vec{B}$ forcing.

## 3. Experimental Setup

Compact and enlarged EMDC configurations were built to probe pressure, compositional, and thermal behavior in a low-field coaxial configuration with the 2T-MHD framework. A low vacuum chamber (LVC) was constructed from 316 stainless steel with an inner diameter of 6" and length of 12". The chamber had vacuum-sealed ConFlat (CF) glass optical windows mounted with diameters of 1.19" to provide enhanced axial visibility. The LVC assembly incorporated DC electrical feedthroughs, Type-T thermocouples, water cooling, and pressure diagnostics. The background pressure within the LVC was measured using an MKS Type AA01A absolute capacitance manometer Baratron connected to an MKS PDR2000 dual-channel gauge controller. The chamber was rough-pumped using an Edwards nXDS15i dry scroll pump at a pumping speed of 15 $m^3$/hr with the pressure being controlled by a Swagelok metering valve. Gas was introduced into the reactor through an upstream inlet, with the flow rate set to 3 SCCM using a Brooks Instrument GF040 mass flow controller.

The gas sampling measurements were performed using 1/16" OD 316 stainless steel capillaries positioned approximately halfway along the reactor axial length. A Swagelok BY Series dual feedthrough transported capillary samples to a high-vacuum chamber (HVC). The HVC was a 36" Dynavac chamber in which gas composition was measured using an MKS e-Vision 2 residual gas analyzer (RGA). It was pumped by an Edwards/CTI Cryo-Torr 400 16" cryopump and a CTI-Cryogenics 9600 water-cooled compressor, with a pumping speed of 5000 L/s for argon.

Two configurations were considered to investigate the influence of thermal boundary conditions and reactor size on EMDC operation. The compact configuration used an inner electrode radius of 2.5 mm and outer electrode radius of 10 mm. This was housed outside the LVC within a quartz glass tube connected to the LVC via an ultra-torr vacuum closure. In this configuration, water and air cooling was applied around the outer tube assembly between the magnets and tube, while the interior discharge region remained uncooled. The enlarged configuration was mounted directly inside the LVC and used an inner electrode radius of 0.6 cm and an outer electrode radius of 3.2 cm (Fig. 7). For this configuration, the water-cooling tubing was coiled around the outer electrode assembly while the interior discharge remained uncooled. In both configurations, the axial fields generated by the NdFeB magnet assembly (ID = 1.5", 3" and OD = 3", 4" for the compact and enlarged configurations respectively) averaged around 0.15 T at the capillary sampling location. The magnets were continuously water-cooled during operation to ensure temperatures were maintained far below the magnet Curie temperature (~583 – 673 K), and the magnetic field strength was monitored using a TD8620 Gauss meter.

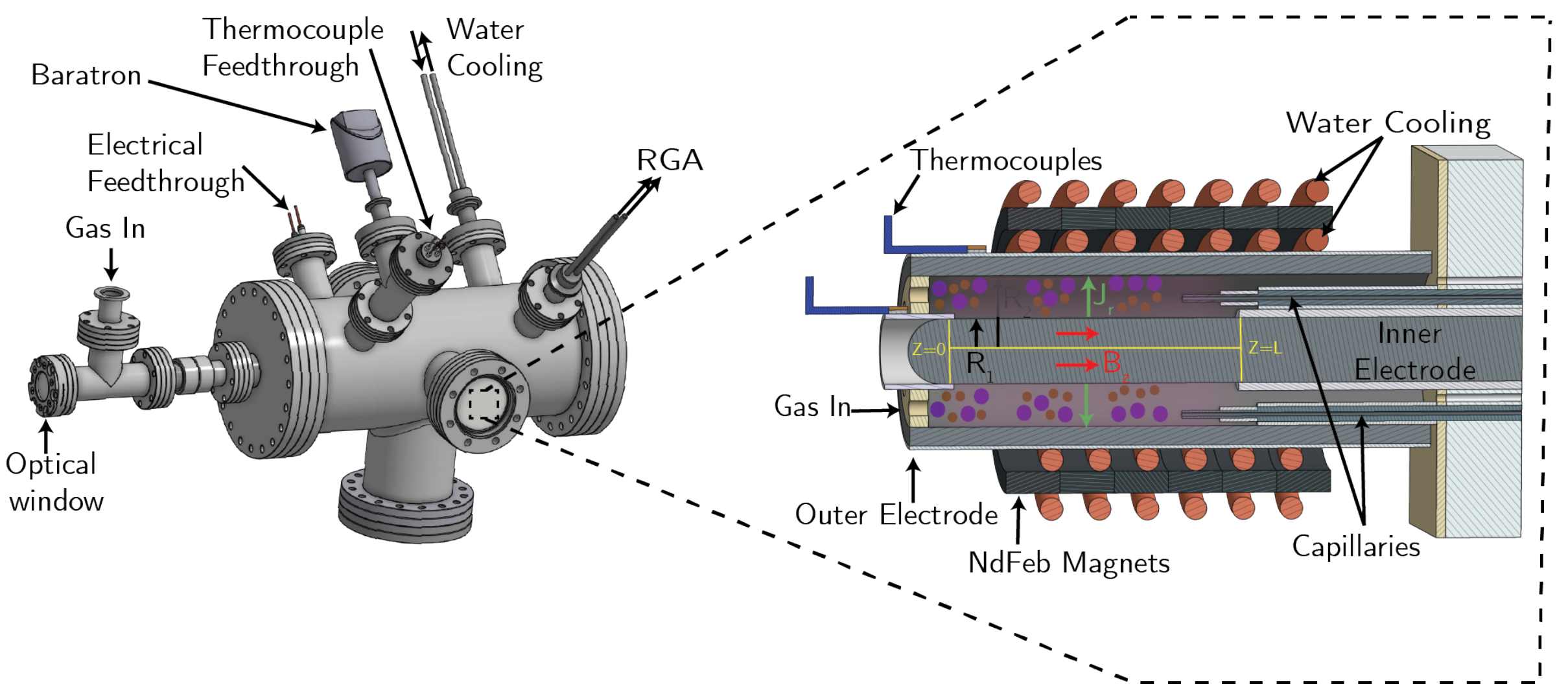

**Figure 7.** EMDC configuration and diagnostic schematic for the enlarged reactor. The left object shows the outside of the LVC with the feedthroughs and attachments labelled. The right schematic shows the coaxial discharge region inside the LVC. The gas is injected into the reactor, and a radial current density is driven between the inner and outer electrodes in the presence of an axial magnetic field generated by NdFeB magnets. Water cooling is provided around the magnet and outer wall assembly, to manage the magnet temperatures during operation. Gas sampling is handled by capillaries routed into the HVC, while thermocouples monitor temperatures along the coaxial boundaries.

The plasma discharge was sustained using a Bdiscom BDS.SP500 DC sputtering power supply. During operation, the currents ranged from 15-250 mA, corresponding to current densities of approximately 31-557 A/m$^2$ depending on the reactor geometry and assumed discharge length. Discharge power ranged from 47 to 80 W. The SP500 could supply an output voltage up to 1 kV with fast arc suppression and recovery mechanisms for stable discharge.

### 3.1. Diagnostics and Measurements

Diagnostics characterized gas composition, temperature, and pressure behavior within the EMDC device. Thermal measurements were taken using thermocouples to monitor the reactor boundary operating temperatures during the plasma discharge. High-speed imaging was used to observe coherent plasma structures under magnetized and unmagnetized discharge conditions [35-38]. Pressure diagnostics established a relationship between the LVC pressure (LVCP) and the sampled capillary measurements through the RGA.

Calibrations were performed to convert the capillary measurements into partial pressures in the HVC. The LVC was filled with a known composition, and the pressure was varied by adjusting the valve to the rough pump. In Fig. 8, this is shown over the experimental range of 2.1-2.5 Torr using a pure argon feed gas (Fig. 8A). At each pressure step, the system was given sufficient time to reach steady state ($dp_{RGA}/dt \sim 0$). This provided a linear relationship between the RGA measurements and the capillaries in the LVC, which were assumed to be equivalent to the Baratron pressure (Fig. 8B). For the 3:1 Ar/Kr mixture in Section 4.1.3, the total measured pressure was partitioned based on the known bottle mixing ratio. The mapping uncertainty was evaluated using standard deviations within the measured data and the covariance values of the linear fitting parameters [39].

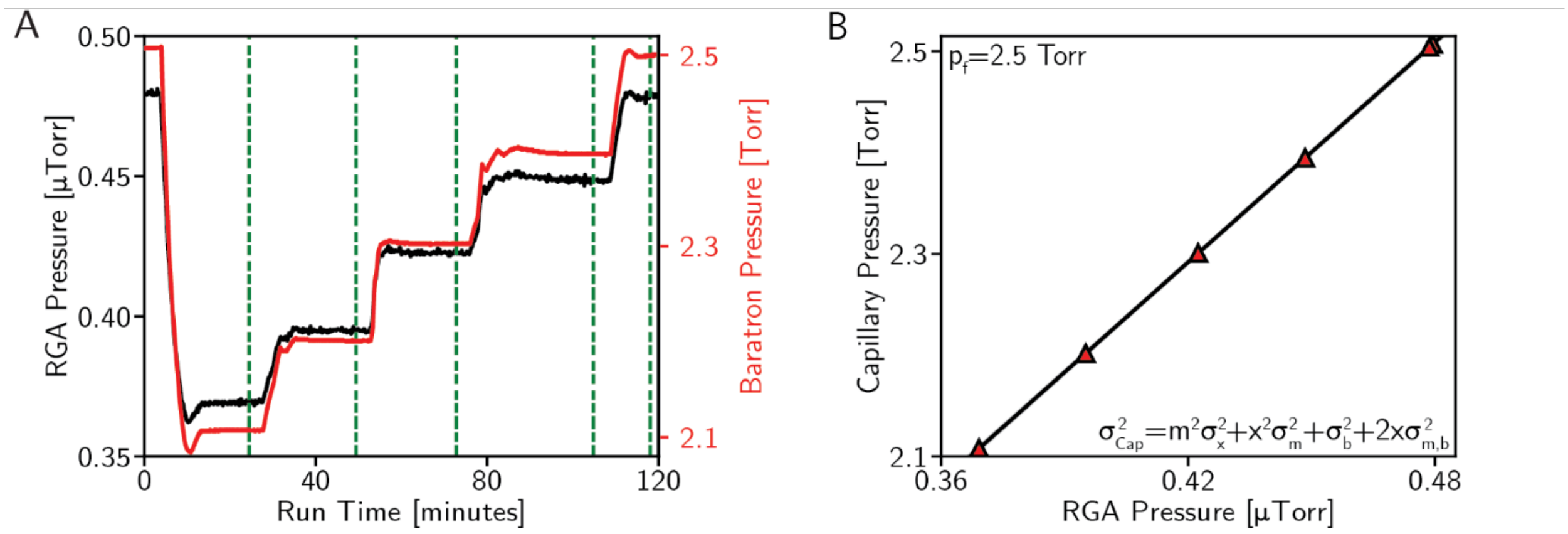

**Figure 8.** Pressure calibration between RGA measurement and LVCP. **(A)** Time-resolved pressure traces during calibration. Both RGA and Baratron pressure measurements are presented in a dual-axis format as the LVCP was stepped ~2.1-2.5 Torr. Dashed vertical lines indicate when the curves were taken to be sufficiently steady state. **(B)** Calibration fit between the RGA pressure and corresponding capillary pressure.

Temperature measurements used thermocouples that recorded temperatures in the enlarged configuration. Thermocouple measurements were recorded using a Keysight 34970A multichannel data acquisition system at a sampling rate of 1 Hz.

High-speed imaging used a Shimadzu HPV-X2 ultra-high-speed video camera with a 10-bit image depth. Images were taken through the axial optical window mounted on the LVC to capture rotating emission structures consistent with spoke behavior reported in cross-field discharges. Images in Fig. 9 were recorded at 1 million frames per second (FPS) with an applied magnetic field of approximately 0.15 T, an initial feed pressure of $p_f \sim 1.5$ Torr, and a discharge current of 100 mA. Without the applied field, ionization waves were not observed during discharge operation (Fig. 9A). The spoke modes were observed to change in number during operation from m = 2 to m = 3 (Figs. 9B and 9C) [40]. The ionization waves rotated azimuthally around the inner electrode cathode as anticipated [41]. Using the frame rate and inner-electrode radius to approximate the velocity of the periodic structures gives phase velocities ($\omega/k$) of approximately 2 km/s (Fig. 9D). However, the phase velocity is not equal to the bulk velocity that drives separation. The pressure measurements presented in Section 4.1.3 indicate the bulk flow velocity was substantially lower ($V_\theta \lesssim 100$ m/s) than the apparent spoke propagation velocity.

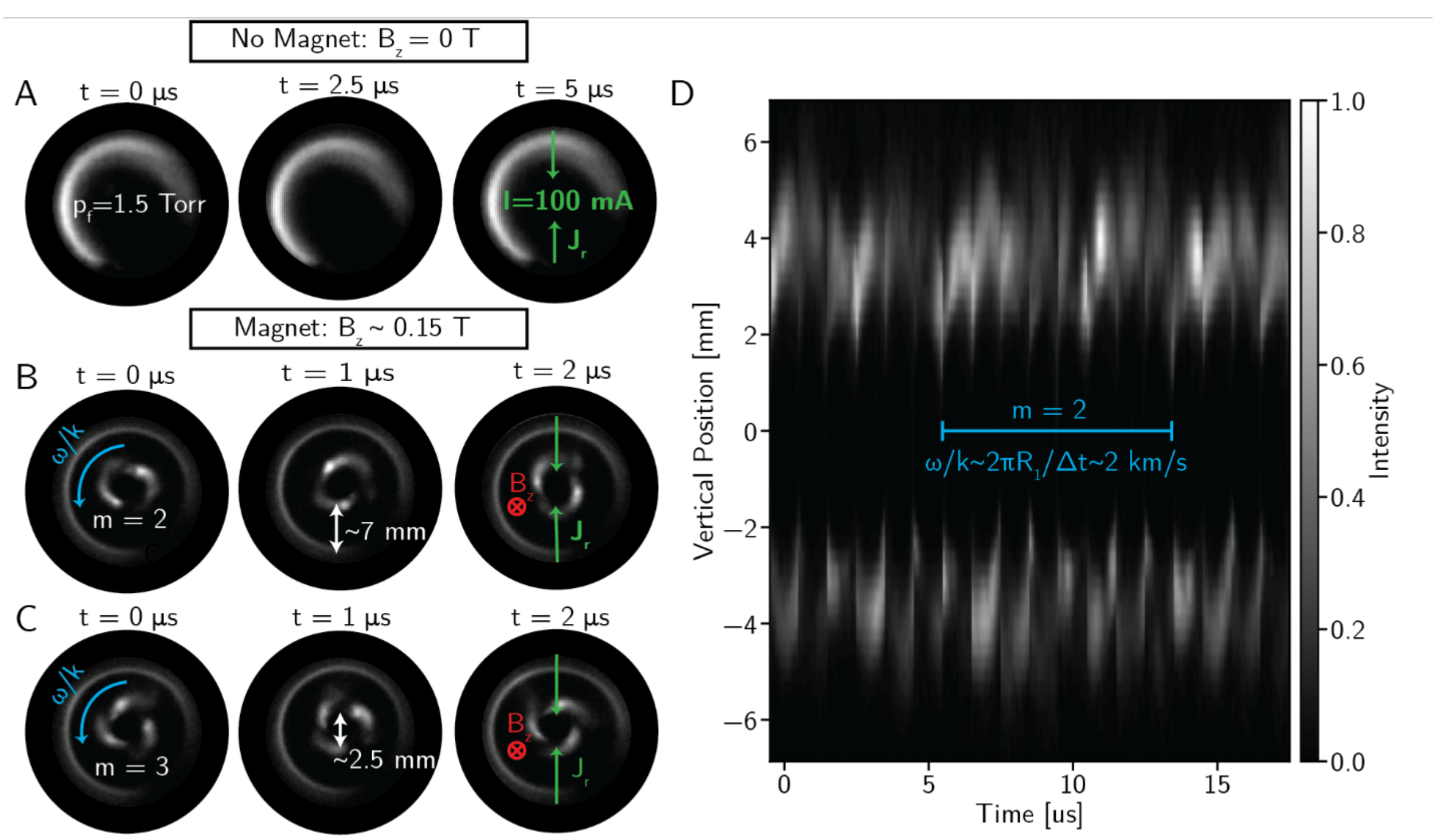

**Figure 9.** High-speed imaging of coherent wave structures **(A)** Image sequence without an applied magnetic field with feed pressure, $p_f$ = 1.5 Torr argon pressure and 100 mA discharge current. **(B,C)** Image sequences with an applied axial magnetic field of B = 0.15 T, showing rotating spoke structures with different azimuthal mode numbers, m = 2 and m = 3. **(D)** Streak-map representation of the emission intensity, showing periodic spoke features. The phase velocity is roughly 2 km/s, but the pressure measurements in Section 4.1.3 support $V_\theta \leq 100$ m/s.

## 4. Results and Discussion

### 4.1 Model Validation

The 2T-MHD model is evaluated against measurements from the literature and the present facility to assess whether it captures the scaling of EMDCs. These include high-current arc discharges and abnormal-glow discharges subjected to high and low magnetic fields, together with low electromagnetic forcing measurements from the present facility. The comparisons emphasize capturing trends in $V_\theta$, $T_h$, p, and $\lambda_h$ rather than an exact reconstruction of each experiment.

Two literature cases are used here to test different routes to electromagnetic forcing: the Wijnakker [8, 9] arc-regime data probe high current density (~15 kA/m$^2$) at low-to-moderate magnetic field (0.13 – 0.26 T), while the Kaneko [42] abnormal-glow data probe high magnetic field (0.57 T) at mild current density (~318 A/m$^2$). Historically, EMDC studies employed analytic models [8, 9, 42] to capture scaling trends. For reference, we make use of these models alongside the 2T-MHD model with rigid-body, $V_\theta \sim Cr$, logarithmic, $V_\theta \sim Cr\ln(r/R_2)$, and Bessel function, $V_\theta \sim CJ_1[kr]$ profiles, where C is a fitting parameter, $J_1$ is the Bessel function of the first kind, and k is the geometric eigenvalue. The logarithmic and Bessel profiles are anchored to satisfy the no-slip boundary conditions. The inner-wall anchored rigid-body profile satisfies the inner-wall condition, but not the outer-wall condition. The shading about the analytic curves represents a $\pm\sigma_x$ standard-error band, where x denotes the fitted parameter for the corresponding analytic profile [39].

The low-field compact and enlarged experiments test the opposite limit by operating in a small geometry ($\Delta R \leq 2.6$ cm), low current densities ($\leq 100$ A/m$^2$), and low magnetic field (~ 0.15 T), where appreciable $\lambda_h$ is not expected. Their purpose was to test the bounds of the model as the geometry, current, and magnetic field strength are reduced.

#### 4.1.1 High-Current Arc Regime

The Wijnakker arc-regime comparison tests whether the 2T-MHD model captures the correct scale of $\lambda_h$ under the highest electromagnetic force conditions in the weakly ionized EMDC literature considered. This case is especially important because it considers the regime most relevant to the model assumptions with high $J_r$, elevated $T_h$, stronger ionization, and ideal radial current flow. The Wijnakker geometry was approximated by assuming a 1D radial current flow in the midplane of their reactor with an effective discharge length of L = 14 cm.

The 2T-MHD predictions were compared with the Wijnakker velocity profiles measured by Doppler shift and with analytical models to see whether each approach recovered the appropriate scaling (Fig. 10). Measured $V_\theta$ profiles at $I = 100$ A for $p_f = 3$ Torr, $B = 0.17$ T (Case 1), and $p_f = 1$ Torr, $B = 0.13$ T (Case 2) are provided across the annulus for an argon feed gas (Fig. 10A). Peak $V_\theta$ of approximately $1100$ m/s for Case 1 and $800$ m/s for Case 2 is predicted by the 2T-MHD model, compared with measured maximum velocities of approximately $950 \pm 135$ m/s and $700 \pm 105$ m/s, respectively. The model and the $V_\theta \sim Cr\ln(r/R_2)$ and $V_\theta \sim CJ_1(kr)$ analytic fits recover comparable peak velocity magnitudes, although the measured profiles are skewed toward the outer wall. This shape difference is likely associated with boundary effects and current-flow structure that is not represented in the EMDC model.

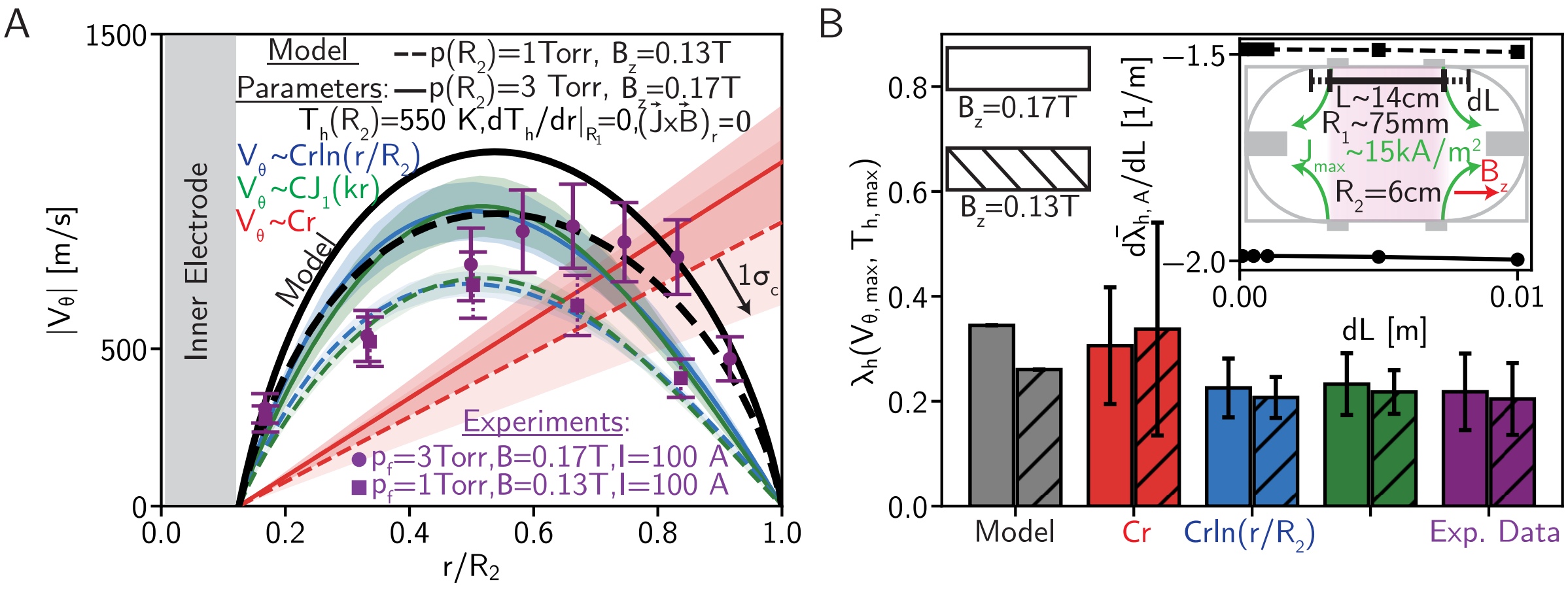

**Figure 10.** Arc-regime comparison between Wijnakker $V_\theta$ measurements, analytic reconstructions, and the EMDC model. **(A)** $V_\theta$ profiles for the reported Wijnakker argon arc cases at 3 Torr, 0.17 T (Case 1) and 1 Torr, 0.13 T (Case 2). Markers show the reported $V_\theta$ Doppler shift measurements with a 15% reproducibility error, while the black curves show EMDC model predictions using the reported operating conditions and an effective $J_r$. The colored curves show the analytic $V_\theta$ fits to reconstruct the $\lambda_h$ values. **(B)** $\lambda_h$ values inferred from the maximum $T_h$ and $V_\theta$ values from the EMDC model, analytic fits, and Wijnakker experimental data for both operating cases. The inset shows the sensitivity of the inferred $\langle \lambda_h \rangle_A$ to the assumed effective discharge length, L = 14 cm.

The inferred $\lambda_h$values in Fig. 10B show that the high-current arc regime reaches the same centrifugal scale in the measurements, analytic reconstructions, and 2T-MHD model. The EMDC model uses the simulated $T_{h,max}$, while the experimental and analytic-fit values use the measured midplane $T_{h,max}$ from the Wijnakker reactor. The $V_\theta \sim C\ln(r/R_2)$ and $V_\theta \sim CJ_1(kr)$ fits produce $\lambda_h$values on par with the experimental estimates of approximately $0.22 \pm 0.07$ for Case 1 and $0.20 \pm 0.07$ for Case 2. The 2T-MHD model predicts somewhat larger values of approximately 0.32 for Case 1 and 0.25 for Case 2.

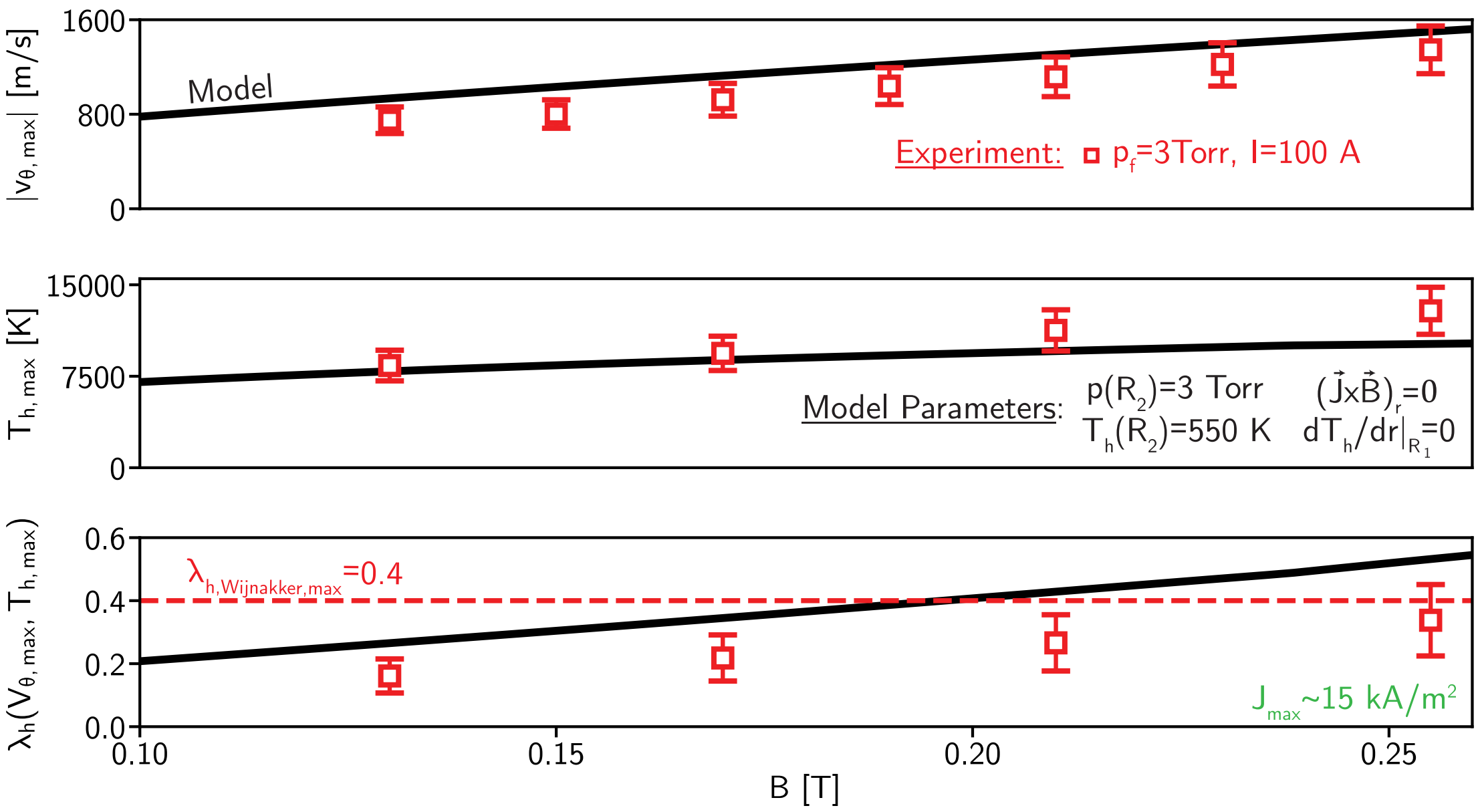

**Figure 11.** Magnetic-field scaling comparison between Wijnakker measurements at $p_f$ = 3 Torr, I = 100 A and EMDC model predictions. The panels compare $V_{\theta,max}$, $T_{h,max}$, and $\lambda_h$ evaluated from the maximum quantities, as functions of the magnetic field. Red markers show Wijnakker experimental values with 15% reproducibility error, while black curves show model predictions. The dashed red line indicates the Wijnakker reported limit of ~ 0.4.

Simulations were also performed with the 2T-MHD model to demonstrate that it can capture the scaling of the arc-regime with increasing magnetic field strengths. Unlike Fig. 10, which compares radial profile shapes for select cases, Fig. 11 isolates the dependence of the characteristic maximum quantities on B for $p_f = 3$ Torr, $I = 100$ A measurements. The simulations were compared against separate experimental measurements of $V_{\theta,max}$, $T_{h,max}$, and $\lambda_h$ from the Wijnakker geometry spanning magnetic fields of 0.1-0.26 T.

The model reproduced the observed increase in $V_{\theta,max}$ with magnetic field and better approached the experimental values as B increases, falling within the 15% reproducibility range by B ~ 0.19 T. The temperature comparison shows the opposite trend, in which the model remains within the correct thermal scale but increasingly underpredicts $T_{h,max}$ at larger B, and it eventually falls outside the reproducibility range at the final experimental point. The resulting $\lambda_h$ values are overpredicted by the model but remain close enough to preserve the experimental scaling. For B = 0.13, 0.17, 0.21, and 0.26T, the experimental estimates of $\lambda_h$ evaluated from $V_{\theta,max}$ and $T_{h,max}$ are approximately $0.21 \pm 0.05$, $0.21 \pm 0.07$, $0.27 \pm 0.09$, and $0.34 \pm 0.11$, while the model predicts approximately 0.26, 0.34, 0.42, and 0.52. The corresponding absolute differences are approximately 0.05, 0.13, 0.15, and 0.18. These residuals show that the model is not an exact reconstruction of the Wijnakker experiment, but it does capture the correct trend of $\lambda_h$ with B over the $J_{max}B \leq 4000$ AT/m$^2$ force range considered in this work.

### 4.1.2 High-Magnetic-Field Abnormal Glow Regime

The Kaneko abnormal glow experiment provides a second validation case from literature that tests the opposite electromagnetic force strategy from the Wijnakker arc case. Whereas Wijnakker used high current density with low-to-moderate magnetic field, Kaneko operated at high magnetic field, $B = 0.57$ T, but comparatively low current density. This comparison is useful because it tests the 2T-MHD model in a regime where magnetization is strong and isolated from the effects of high current densities.

Simulations with the 2T-MHD model were performed with argon for $p_f = 0.5$ Torr, $I = 0.5$ A, and $B_z = 0.57$ T (Fig. 12A). Unlike the Wijnakker data, where velocity and temperature measurements can be compared directly, the Kaneko comparison reconstructs $V_\theta^2/T_h$ from the measured radial pressure ratio. The analytic profiles are fitted in log-pressure space, $\ln(p/p(r_1))$, with the innermost pressure measurement used as the normalization point. For the 2T-MHD calculation, the effective discharge length was taken as approximately 5 cm based on the reported axial length of the inner-electrode ring region where the radial current was concentrated.

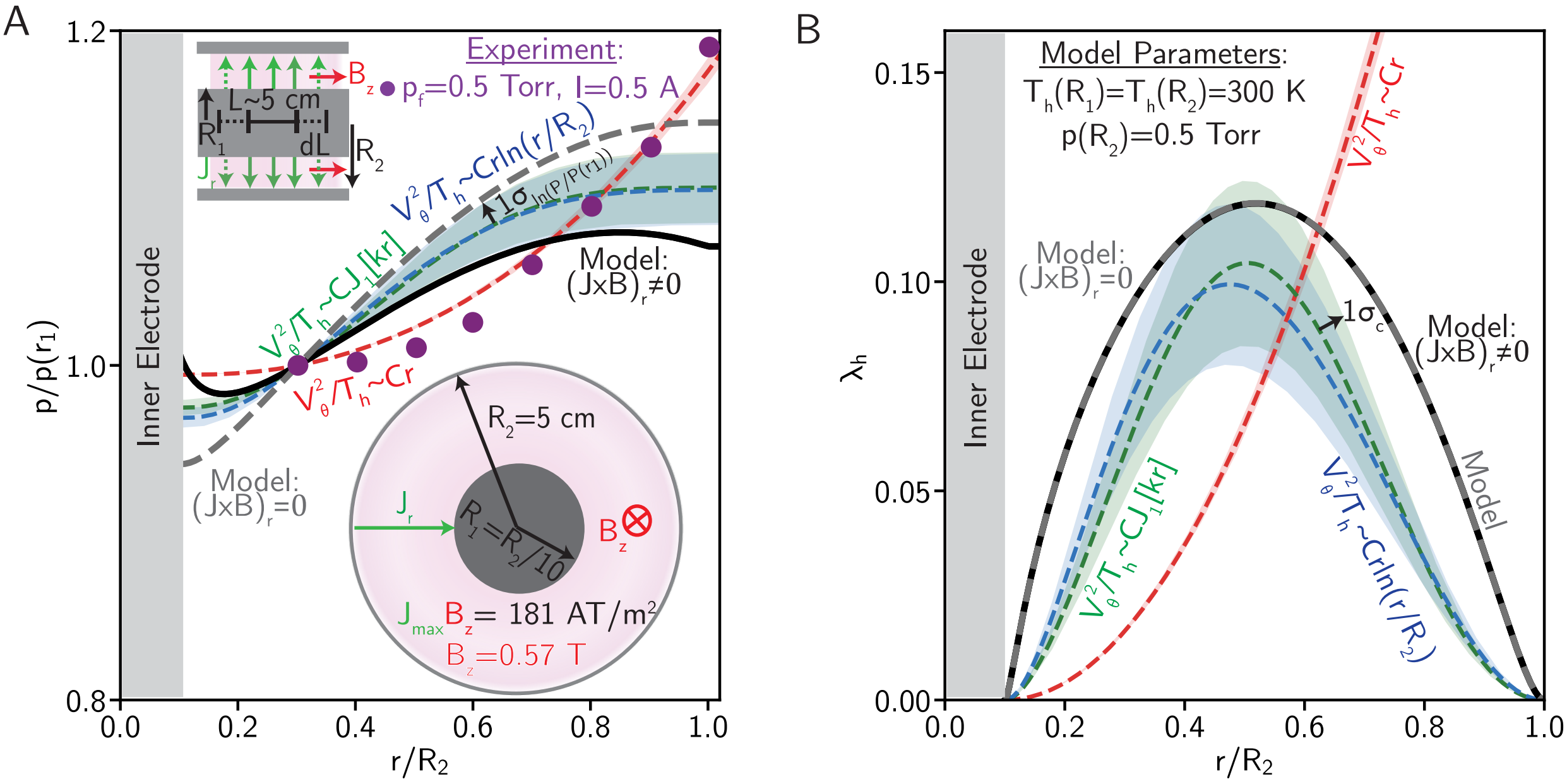

**Figure 12.** High magnetic-field, low current density abnormal glow validation. **(A)** Radial pressure ratio, $p/p(r_1)$, profile for the reported measurements using B = 0.57 T and estimated electromagnetic force $J_{max}B_z = 181$ AT/m$^2$. The black and grey curves show the 2T-MHD model with the $(\vec{J} \times \vec{B})_r$ term activated and deactivated. Colored curves show pressure profiles reconstructed from isothermal analytic $V_\theta^2/T_h$ fits. The shaded bands indicate the propagated standard error in $\ln(p/p(r_1))$. **(B)** Corresponding $\lambda_h$ profiles inferred from the pressure reconstructions and the 2T-MHD model.

The rigid-body reconstruction gives the closest match to the measured pressure profile in Fig. 12A. This is consistent with the pressure-profile behavior also observed in reported Wijnakker data [8, 9]. However, the agreement should not be interpreted as proof that the flow is rigid body. The rigid-body profile does not satisfy the outer-wall no-slip condition and produces a $V_\theta$ and $\lambda_h$ structure that differs strongly from the other profiles in Fig. 12B. The 2T-MHD model underpredicts the outer-wall pressure rise but reproduces the correct order of magnitude when $(\vec{J} \times \vec{B})_r = 0$. The measured value is approximately $p(R_2)/p(r_1) \sim 1.19$. The 2T-MHD case with $(\vec{J} \times \vec{B})_r = 0$ predicts $p(R_2)/p(r_1) \sim 1.15$, while the logarithmic and Bessel reconstructions give approximately $1.11 \pm 0.02$. Including $(\vec{J} \times \vec{B})_r$ produces a larger radial pressure shift that appears to be overestimated for this experiment, likely because the reduced 1D model does not resolve the actual two-dimensional current and pressure distribution. This interpretation is supported by the fact that the Bessel and logarithmic fits produce similar profiles to the 2T-MHD model, and they all share common 1D steady-state laminar assumptions.

The pressure profiles are converted into inferred $\lambda_h$ profiles (Fig. 12B). The 2T-MHD model predicts a peak value near $\lambda_h \sim 0.12$, while the non-rigid analytic reconstructions give peak values near $\lambda_h \sim 0.10 \pm 0.02$. Toggling $(\vec{J} \times \vec{B})_r$ has little effect on the inferred $\lambda_h$ profile, even though it noticeably shifts the pressure profile in Fig. 12A, as previously indicated in Fig. 3. Overall, the Kaneko comparison supports the use of the 2T-MHD model for estimating the $\lambda_h$ scale in a high-B, low-J abnormal-glow regime, while showing that detailed pressure-shape agreement requires physics outside the present 1D formulation.

**4.1.3 Low-Field Compact and Enlarged Reactor Regimes**

Experiments were performed at UT-Austin to test the scaling of EMDCs at low currents, smaller geometries, and with binary mixtures having large component mass differences. The purpose of these comparisons is not to demonstrate large $\lambda_h$ or strong separation, but to test whether the 2T-MHD model is appropriately bounded when $J_{max}B$ and $\Delta R$ are small.

The compact reactor used pure Ar and two capillary sampling locations near the inner and outer radii, $r_1 \sim 5$ mm and $r_2 \sim 8$ mm. Pressure measurements showed that the inner and outer capillary signals decreased linearly with increasing current (Fig. 13A). This produced little resolvable radial dependence between the two sampling locations. The observed behavior is consistent with a low-force regime where the dominant pressure change is associated with bulk gas heating and axial expansion rather than centrifugal redistribution.

The measured steady-state pressure ratio between the inner and outer capillaries in Fig. 13A was compared with the EMDC model (Fig. 13B). The experimental ratio remains near unity within uncertainty, and the model predicts the same weak radial pressure response over the measured $J_{max}B$ range. The extended model curve indicates that a noticeable pressure-ratio shift in this compact geometry would require $J_{max}B$ values approaching $10^3$ AT/m$^2$. Therefore, the compact reactor geometry and applied force are insufficient to produce a measurable centrifugal pressure shift by the RGA, consistent with the small $\Delta R$ and low $J_{max}B$ scaling predicted in Fig. 5.

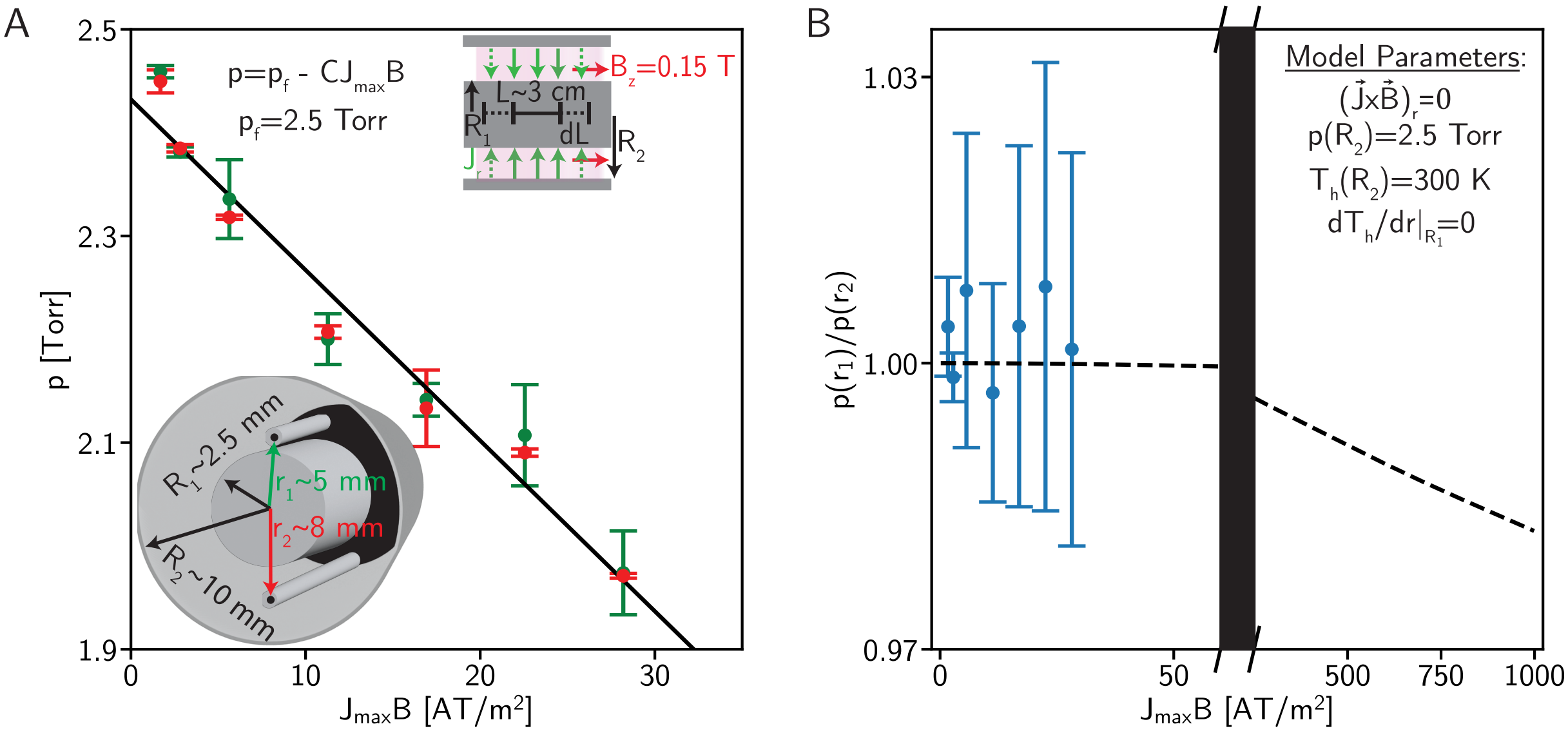

**Figure 13.** Low-field compact capillary pressure measurements. **(A)** Absolute capillary pressure measured at the inner and outer radial sampling positions as a function of electromagnetic forcing, $J_{max}B$. Green and red markers denote the inner and outer capillary sampling positions, respectively. The pressure decreases linearly with electromagnetic forcing and is represented using the empirical relation $p = p_f - CJ_{max}B$, where C is a fitted coefficient. **(B)** Measured radial pressure ratio, $p(r_1)/p(r_2)$, compared with the EMDC model. The pressure ratio in the left portion of panel B remains close to 1 over the applied $J_{max}B$ range, consistent with the prediction that the compact geometry and low forcing produce negligible radial pressure gradients. The right portion illustrates that an appreciable pressure ratio reduction requires substantially larger $J_{max}B$, although the predicted shift remains small for the compact reactor geometry.

Low-field temperature scaling was evaluated using the 2T-MHD model for an argon feed gas in the enlarged reactor. Fig. 14A shows that the temperature measured by the inner thermocouple rose substantially more than the temperature measured by the outer thermocouple during operation at $p_f \sim 2.5$ Torr, $I = 25$ mA, and $B = 0.15$ T. The measured steady-state wall temperatures are approximately 340 K at the inner location and 300 K at the outer location, corresponding to $\Delta T \sim 40$ K. Fig. 14B demonstrates that $\Delta T_h$ increases with current and that the EMDC model captures the correct trend, although it still underpredicts the measured temperature difference. We speculate that this difference is caused by the assumption of Saha equilibrium, which overpredicts electrical conductivity and thus reduces $\vec{J} \cdot \vec{E}^*$ in nonequilibrium conditions [30].

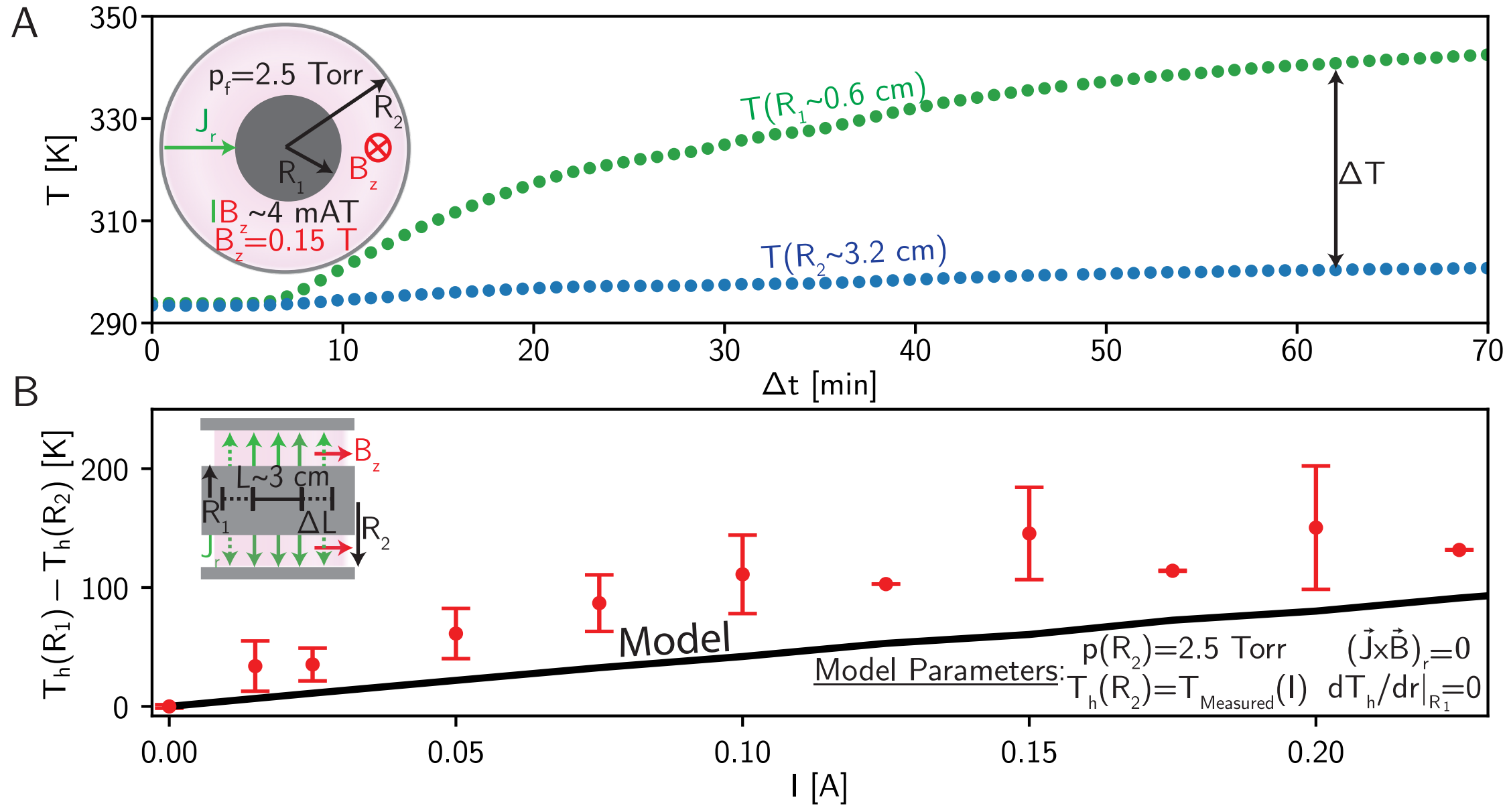

**Figure 14.** Thermocouple validation measurements in the enlarged reactor regime for argon. **(A)** Thermocouple measurements at the inner and outer radial positions during reactor operation ($P_f$~2.5Torr, I = 25 mA, and B~0.15T). **(B)** Measured radial temperature difference, $\Delta T_h = T_h(R_1) - T_h(R_2)$, as a function of discharge current compared with the EMDC model prediction. Experimental values are shown as markers with uncertainty bars arising from the standard deviation of the steady-state temperature measurements.

The enlarged reactor was then used to increase the radial spacing between capillaries and test whether a larger $R_2$ generates a measurable compositional shift during operation. The capillaries sampled an outer location near $r_2 \sim 3$ cm and an inner location near $r_1 \sim 1$ cm. Fig. 15A shows calibrated Ar and Kr partial-pressure traces during operation. After plasma ignition, both partial pressures decreased as expected from Fig. 13A. When the outer-capillary signal reached a steady condition, the sampling line was switched to the inner capillary.

The mole-fraction ratios with an applied I = 100 mA in Fig. 15B show the expected trend for mass-dependent radial separation. The lighter Ar fraction increases at the inner capillary relative to the outer capillary, and the heavier Kr fraction decreases. The 2T-MHD model is not appropriately scoped to capture the scaling of a non-dilute Ar/Kr mixture, and the measured composition shift is larger than expected from the 2T-MHD model under the same low forcing conditions. Using the argon transport closures and an assumed effective discharge length of L ~ 3 cm, the model predicts an outer-capillary velocity of only $|V_\theta(r_2)| \sim 1$ m/s and a negligible composition shift. In contrast, reproducing the measured capillary-to-capillary shift with an isothermal rigid-body estimate using $\bar{T}_h \sim 377$ K from Fig. 14 requires $|V_\theta(r_2)| \sim 75$ m/s. We attribute these differences partly to current filamentation caused by the geometry of our reactor. In the configuration studied, axial heating during the discharge caused the pressure to rise, the plasma to narrow, and thus J to increase as the discharge developed.

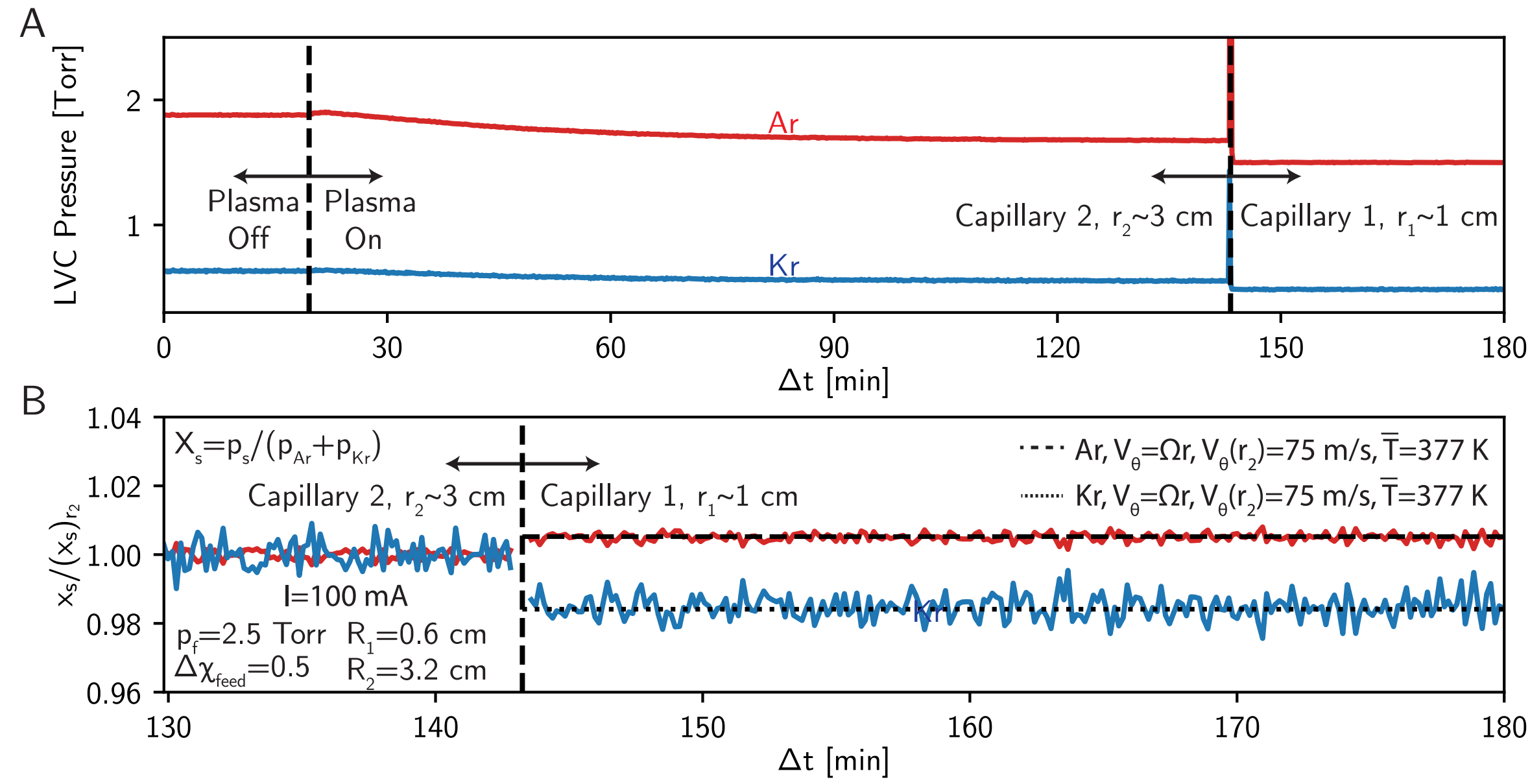

**Figure 15.** Capillary traces for the enlarged reactor during a 3:1 two-component Ar/Kr discharge. **(A)** Calibrated LVC partial-pressure traces for argon and krypton displaying reactor ignition and capillary switching. The signal is first sampled at the outer capillary location, ($r_2 \sim 3$ cm), and then switched to the inner capillary ($r_1 \sim 1$ cm). **(B)** Outer capillary normalized mole-fractions derived from the partial-pressure traces. After switching to the inner capillary, $\chi_{Kr}/(\chi_{Kr})_{r_2}$ decreases and $\chi_{Ar}/(\chi_{Ar})_{r_2}$ increases. The dashed and dotted lines are the overlaid isothermal rigid-body model predictions for argon and krypton, respectively, for an outer-capillary velocity, $|V_\theta(r_2)| \sim 75$ m/s and $\bar{T} \sim 377$ K. This capillary-to-capillary shift is qualitatively consistent with weak centrifugal separation, although its magnitude may be amplified by current filamentation and the evolving discharge geometry.

## 4.2 $\lambda_h$ Scaling with Current and Magnetic Field

With the 2T-MHD model demonstrated to capture scaling trends of $\lambda_h$, the model was then used to extrapolate EMDC performance with electromagnetic force parameters $J_{max}$ and B. The purpose of this sweep is to determine whether $\lambda_h \sim 1$ can be reached within a literature-informed operating envelope, and whether it is more effective to increase current density or magnetic field. The geometry is fixed at a favorable annular condition informed by Fig. 5 ($R_2 = 6$ cm, $R_2/R_1 = 3$), while $J_{max}$ is varied from $10^3$ to $10^4$ A/m$^2$ and B from 0.1 to 0.4 T. Here, we are focusing on the regime where $J_{max}$ is high, which is where the 2T-MHD model was found to be most accurate during validation.

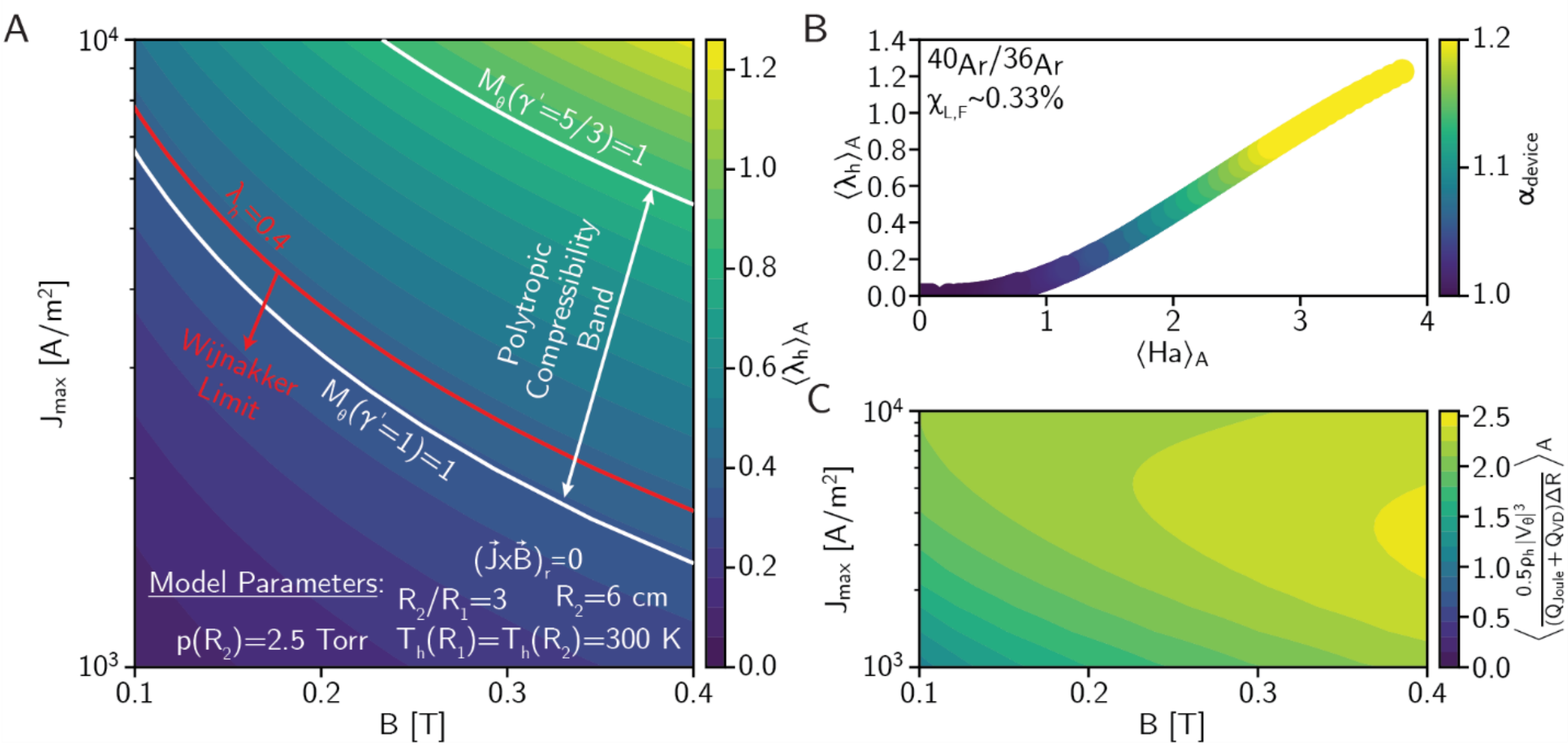

**Figure 16.** $\langle\lambda_h\rangle_A$, $\langle Ha\rangle_A$, $\alpha_{device}$, and power-efficiency scaling with $J_{max}$ and B in the arc-regime design space for $R_2 = 6$ cm and $R_2/R_1 = 3$. **(A)** Area-averaged $\lambda_h$ as a function of $J_{max}$ and B. The red contour marks the reported Wijnakker upper limit value, $\lambda_h \sim 0.4$, and the white contours indicate approximate $M_\theta(\gamma') = 1$ compressibility limits for isothermal, $\gamma' = 1$, and isentropic, $\gamma' = 5/3$, pressure responses. **(B)** Relationship among $\langle\lambda_h\rangle_A$, $\langle Ha\rangle_A$, and $\alpha_{device}$ for a $^{40}Ar/^{36}Ar$ mixture with $\chi_{L,F} \sim 0.33\%$. **(C)** Scaling of the nondimensional efficiency metric $\langle 0.5\rho_h \mid V_{\theta h} \mid^3/((Q_{Joule} + Q_{VD})\Delta R)\rangle_A$. This demonstrates how the preferential conversion of electromagnetic forcing into rotational energy flux rather than heating depends on the partition between $J_{max}$ and B.

The maximum value in the displayed range is approximately $\langle\lambda_h\rangle_A \sim 1.2$ near $J_{max} = 10^4$ A/m² and $B = 0.4$ T (Fig. 16A). For reference, the reported Wijnakker upper value of $\lambda_h \sim 0.4$ is included as a contour. The contour plot indicates that Wijnakker centrifugal conditions can be reached over a broad range of $J_{max}$ and B combinations, but values approaching $\lambda_h \sim 1$ require simultaneous increases in electromagnetic force and favorable geometry. This result suggests that the $\lambda_h < 1$ limitation is not a universal bound. In the present model, the attainable $\lambda_h$ depends on how Lorentz forcing, Joule heating, magnetized transport, and geometry are coupled.

Because the maximization of $\lambda_h$ involves optimizing $V_\theta^2/T_h$, compressibility effects could become relevant as $V_\theta$ scaling outpaces $T_h$. To estimate where this could occur, the white contours marking $M_\theta = 1$ in Fig. 16A indicate where compressibility may become important. A reduced pressure closure [20] is used to parameterize the pressure-density response about the 2T-MHD solution, and the $\gamma' = 1$ and $\gamma' = 5/3$ contours correspond to isothermal and isentropic pressure perturbation limits. The region between these contours is interpreted as a polytropic compressibility band, indicating where compressibility and the possibility of wave or shock formation may require further consideration. However, it should be stressed that SDCs frequently operate above sonic speeds, so this is not a unique feature of EMDCs.

The electromagnetic force swept in Fig. 16A is mapped to $\alpha_{device}$ by plotting the corresponding $\langle\lambda_h\rangle_A$ against $\langle Ha\rangle_A$, with the individual points colored by their associated $\alpha_{device}$ (Fig. 16B). The highest forcing conditions give $\langle\lambda_h\rangle_A \sim 1.2$, $\langle Ha\rangle_A \sim 3.8$, and $\alpha_{device} \sim 1.2$. As expected, the larger $\alpha_{device}$ values occur when elevated $\lambda_h$ is sustained with sufficient Ha coupling, so that the integrated $\lambda_h/r$ value is greater.

The corresponding scaling of the nondimensional energy flux efficiency metric, $\langle 0.5\rho_h \mid V_\theta \mid^3/((Q_{Joule} + Q_{VD})\Delta R)\rangle_A$ compares the $J_{max}$ and B force scaling of rotational energy relative to the thermal dissipation channels from Joule heating ($Q_{Joule}$) and viscous dissipation ($Q_{VD}$) (Fig. 16C). The results indicate that $J_{max}B$ is not sufficient to characterize EMDC efficiency. Although the total product sets the scale of the Lorentz forcing, its partition between $J_{max}$ and B affects magnetization, $Q_{Joule}$, ionization, and ultimately the evolution of $\langle\lambda_h\rangle_A$ and $\langle Ha\rangle_A$. Increasing B can be more favorable than increasing $J_{max}$ alone because it strengthens volumetric force coupling while reducing the current density required to reach a fixed force. This is consistent with broader electromagnetic-flow efficiency scaling in magnetized plasma devices [43]. However, the high-efficiency region narrows as B increases. This occurs because stronger magnetization also reduces $\sigma_P$, increasing the Joule-heating penalty associated with radial current.

Together, Fig. 16 shows that EMDC performance is controlled by coupled optimization, not by maximizing $J_{max}B$ alone. Larger B can improve the conversion of electromagnetic forcing into rotation, but only over a narrower range of compatible $J_{max}$. Higher $J_{max}$ increases forcing but also increases Joule heating through $\vec{J} \cdot \vec{E}^* \sim J_r^2/\sigma_P$. The most favorable regime is therefore one in which B, $J_{max}$, and the geometry are chosen so that the increase in $V_\theta$ sufficiently outpaces the increase in $T_h$. Under these conditions, the model predicts that $\langle\lambda_h\rangle_A \geq 1$ is attainable.

### 4.3 Optimization of $^{40}Ar/^{36}Ar$ Separation

Expanding upon the trends from Section 4.2, Fig. 17 evaluates how radial geometry changes the separative performance of an EMDC under optimized field conditions and compares the result with SDC predictions. The curves use a $^{40}Ar/^{36}Ar$ mixture with $\chi_{L,F} \sim 0.33\%$, an outer radius $R_2 = 6$ cm, $B_z = 0.45$ T, and $J_{max} = 7.5$ kA/m$^2$. The SDC reference is evaluated at a representative carbon-fiber thin-wall speed limit of $V_W = 720$ m/s.

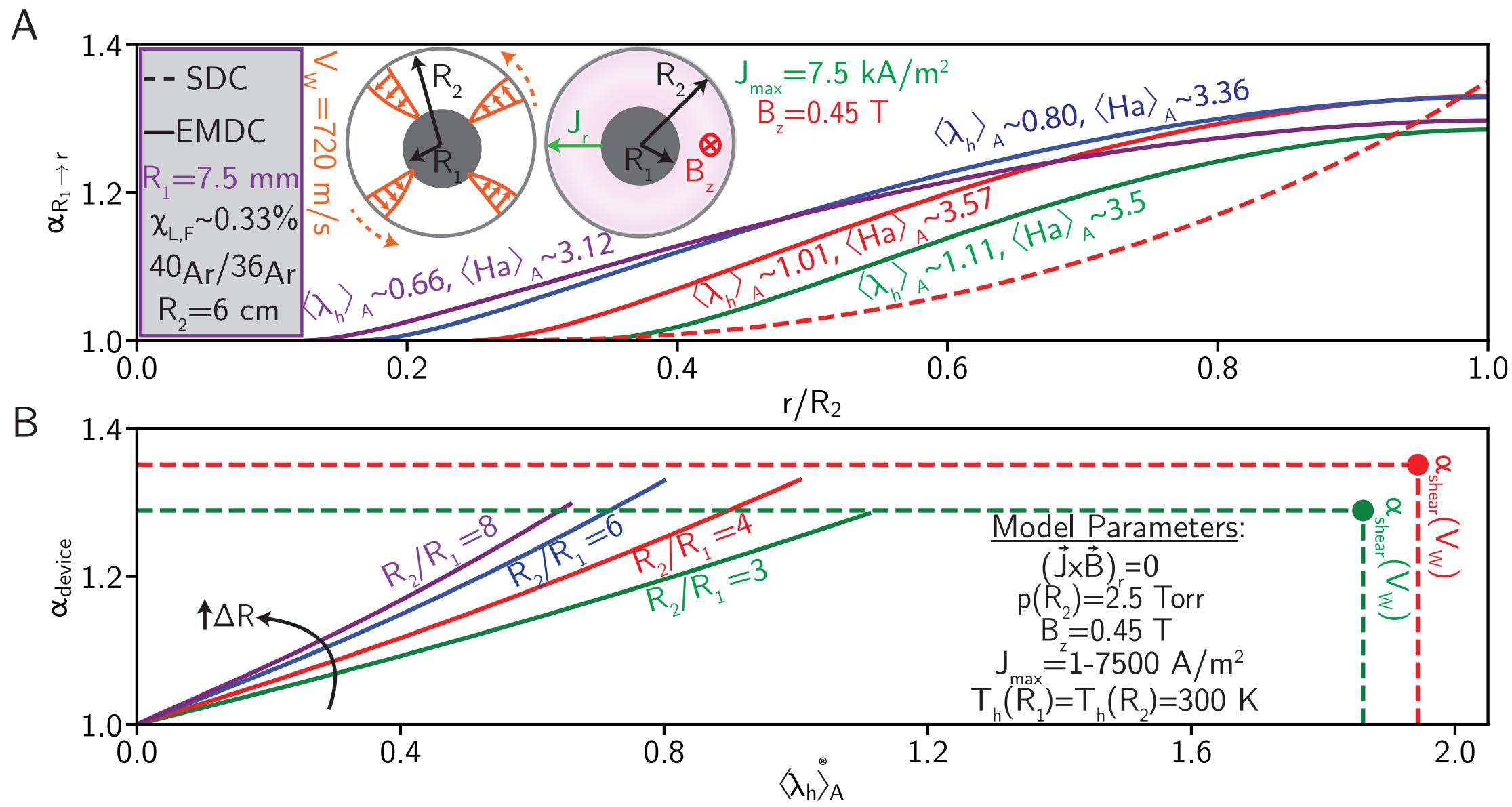

**Figure 17.** $^{40}Ar/^{36}Ar$ separation-factor scaling for EMDC radial geometries compared with SDC predictions at the carbon-fiber thin-wall speed limit of $V_W = 720$ m/s. The EMDC cases use $R_2 = 6$ cm and $B_z = 0.45$ T. **(A)** $\alpha_{R_1\to r}$ profiles for different $R_2/R_1$ values. Solid curves show EMDC predictions with $J_{max} = 7.5$ kA/m$^2$ and the red dashed curve shows SDC predictions for $R_2/R_1 = 4$. **(B)** $\alpha_{device}$, as a function of $\langle\lambda_h\rangle_A$, obtained by sweeping $J_{max}$ at fixed $B_z = 0.45$ T. The results show that EMDC performance depends on the coupled effects of $\langle\lambda_h\rangle_A$, $\langle Ha\rangle_A$, and $\Delta R$. The $R_2/R_1 = 3$ EMDC case matches the corresponding SDC $\alpha_{device}$ at lower $\langle\lambda_h\rangle_A$, while the red and blue EMDC cases approach but remain below the $R_2/R_1 = 4$ SDC device-level benchmark within the range considered. The EMDC profiles nevertheless exceed the SDC $\alpha_{R_1\to r}$ over substantial portions of the annulus.

Radial separation is controlled by the integrated effect of $\lambda_h/r$, not by $\langle\lambda_h\rangle_A$ alone as indicated by Fig. 17A. The blue EMDC case, with $R_2/R_1 = 6$, produces a larger $\alpha_{R_1\to r}$ than the red and green cases despite having lower $\langle\lambda_h\rangle_A$ and $\langle Ha\rangle_A$. This occurs because the larger radial integration length, $\Delta R$, provides a longer interval over which $\lambda_h/r$ is integrated. However, this trend is not monotonic. The purple case, $R_2/R_1 = 8$, performs worse than the blue and red cases because its geometry reduces the ability of the imposed electromagnetic force to sustain favorable $\langle\lambda_h\rangle_A$ and $\langle Ha\rangle_A$. Thus, increasing $\Delta R$ can offset reduced $\lambda_h/r$ values, but $\alpha_{R_1\to r}$ remains dictated by the integrated profile.

Comparing $\alpha_{device}$ against $\langle\lambda_h\rangle_A$ for the same radial geometries shows that the EMDC cases produce greater $\alpha_{device}$ per $\langle\lambda_h\rangle_A$ (Fig. 17B). The green $R_2/R_1 = 3$ EMDC case reaches the corresponding SDC $\alpha_{device}$ at substantially lower $\langle\lambda_h\rangle_A$. The red and blue EMDC cases approach the larger $R_2/R_1 = 4$ SDC $\alpha_{device}$ but remain slightly below within the sweep.

The main implication of Fig. 17 is not that every EMDC geometry exceeds the $\alpha_{device}$ of an SDC. Rather, volumetric forcing provides an additional means of shaping the radial distribution of $\lambda_h$. The EMDC advantage comes from the ability to tune $\vec{J} \times \vec{B}$ and radial geometry to shape $\lambda_h$, set the integration length $\Delta R$, and produce favorable values of $\langle \lambda_h \rangle_A$ and $\langle Ha \rangle_A$, thereby maximizing the integral of $\lambda_h / r$. When the electromagnetic force and geometric conditions are jointly optimized for this purpose, the 2T-MHD model predicts that an EMDC can exceed the SDC $\alpha_{R_1 \to r}$ over much of the annulus and, for a favorable geometry, match the corresponding SDC $\alpha_{device}$ while requiring a lower $\langle \lambda_h \rangle_A$.

## 5. Summary and Conclusions

This work examined whether EMDCs can convert $\vec{J} \times \vec{B}$ forcing into competitive radial mass separation before thermal dissipation eliminates their advantage of higher attainable $V_\theta$. The main result is that $\lambda_h$ alone is insufficient to capture centrifugal performance scaling. Device-level separation depends on the radial integral of $\lambda_h / r$, so $\Delta R$, the radial distribution of $\lambda_h$, and the momentum balancing mechanisms quantified by Ha are also central to determining $\alpha_{device}$.

A 2T-MHD model was developed to connect the operating inputs $J_r$, $B_z$, $R_2$, $R_2/R_1$, and $p(R_2)$ to the coupled profiles of $T_h$, $T_e$, p, and $V_\theta$. The model shows that the dominant thermal penalty is not $\overleftrightarrow{\tau} : \nabla\vec{V}$. Instead, $\vec{J} \cdot \vec{E}^*$ is amplified by the magnetization-induced suppression of the Pedersen conductivity, $\sigma_P$. This result revises prior weakly ionized plasma centrifuge results that treated $\lambda_h < 1$ as a universal consequence of $V_\theta^2$ not being able to scale faster than $T_h$ with increasing $\vec{J} \times \vec{B}$ due to $\overleftrightarrow{\tau} : \nabla\vec{V}$.

Geometric studies demonstrated that EMDC performance is controlled by how momentum is balanced through the annulus in steady state. Larger radii and intermediate aspect ratios are favorable because they reduce wall-loss dominance while being able to effectively concentrate the imposed current in the gap. As a result, the largest $\alpha_{device}$ and $\alpha_{R_1 \to r}$ do not occur in the case with the largest peak $\lambda_h$ or $\langle \lambda_h \rangle_A$. A lower $\lambda_h$ case can outperform a higher $\lambda_h$ case if the integrated $\lambda_h / r$ profile is higher, which is attributed to $\langle \lambda_h \rangle_A$, $\langle Ha \rangle_A$, and $\Delta R$.

The optimized $^{40}Ar/^{36}Ar$ calculations demonstrate the main separative implication. EMDCs do not need to reproduce the high wall-localized $\lambda_h$ of SDCs to obtain comparable $\alpha_{device}$ or higher $\alpha_{R_1 \to r}$. Because the EMDC introduces momentum volumetrically rather than through a rotating wall, moderate $\lambda_h$ can be maintained over a larger fraction of the annulus and still be competitive with SDCs. By tuning $\vec{J} \times \vec{B}$ and radial geometry, the EMDC can exceed the SDC $\alpha_{R_1 \to r}$ over much of the annulus and, in the favorable geometry demonstrated here, match the corresponding $\alpha_{device}$ at lower $\langle \lambda_h \rangle_A$. The relevant operating and design variables are therefore $J_{max}$, $B_z$, and radial geometry. The resulting Ha and $\lambda_h$ distributions provide diagnostic metrics for evaluating the optimized state.

EMDCs and SDCs should not be compared solely using peak or area-averaged $\lambda_h$. The EMDC advantage comes from volumetric electromagnetic forcing and additional transport tunability through geometry-dependent momentum redistribution mechanisms that promote a broader annular $\lambda_h$ profile. Additionally, the heating within EMDCs lends itself to operating on feedstocks that must be thermally treated, and the volumetric forcing that produces higher $\alpha_{R_1 \to r}$ is especially favorable for high-$\Delta m$, high-throughput applications. Further work is required to determine whether the favorable one-dimensional scaling is retained when finite-length effects, secondary flows, MHD instabilities, and cascade-level engineering constraints are considered. The present study therefore reframes the EMDC design problem from maximizing $\lambda_h$ and shifts the emphasis to its radial distribution, together with Ha and $\Delta R$, to exceed the SDC $\alpha_{R_1 \to r}$ over substantial portions of the annulus and, for the favorable geometry demonstrated here, match the corresponding SDC $\alpha_{device}$ at lower $\langle \lambda_h \rangle_A$.

**Acknowledgements**

The authors wish to acknowledge the assistance of Harsha Rajesh, Mark Dunn, and Georgy Krasovskiy in building the experimental facility and taking measurements. The authors also wish to acknowledge Georgy Krasovskiy and Marco Obregon for their contributions to the CAD design of experimental components used in this work.

**Author Contributions**

Drue Hood-McFadden performed the experiments, developed the computational model, analyzed the data, undertook the literature review, conceptualized the narrative, and drafted the paper text and figures. Shreyas Kotla aided in performing the experiments, reviewing the literature, and drafting the text and figures. Thomas C. Underwood was responsible for the initial conceptualization and funding of the project, and for assisting in the narrative direction.

**Funding**

The authors wish to acknowledge funding support through the University of Texas at Austin Energy Institute Fueling the Sustainable Energy Transition (FSET) program.

**Appendix A: Reduced 2T-MHD Formulation Implemented in Numerical Model**

The 2T-MHD formulation implemented in the numerical model treats the EMDC as composed of $\mathcal{S} = \{e, Ar, Ar^{+}\}$, where $\bar{m}_h \sim m_{Ar}$. The imposed force is the radial current density $J_r = \pm J_{max} R_1 / r$ in an externally applied axial magnetic field $B_z$, which produces the azimuthal Lorentz force $(\vec{J} \times \vec{B})_\theta = -J_r B_z$. The azimuthal momentum equation balances this volumetric electromagnetic forcing against viscous shear stress $(d(r^3 \mu_n d(V_\theta/r)/dr))/r^2 dr)$. The radial pressure equation includes both a centrifugal term, $\rho_h V_\theta^2/r$, and the radial Lorentz force contribution, $J_\theta B_z$. The heavy-particle energy equation balances radial thermal conduction $(d(r\kappa_n dT_h/dr)/rdr)$, ion Joule heating $(J_{ri} E_r^*)$, and viscous dissipation $(\mu_n (rd(V_\theta/r)dr)^2)$ against electron-heavy-particle energy exchange $(3 n_e k_B m_e \nu_m^e (T_h - T_e)/m_n)$. From Eq. 5 with the simplifications to $\vec{\xi}_s$, the fluid frame electric field is $E_r^* = (J_r - \sigma_{pi} \varepsilon_{ri}^*)/\sigma_p$, where $\varepsilon_{ri}^* = m_i V_\theta^2/er$. Finally, the electron energy relation is treated algebraically through local electron Joule heating $(\sigma_{pe} (E_r^*)^2)$ and elastic energy transfer to the heavy gas.

The charged species are not treated as separate momentum fluids. Instead, their drift response is closed algebraically through the magnetized conductivity tensor, with $\sigma_p$ and $\sigma_H$ components evaluated from the local collision frequencies $(\nu_m^{s_q})$, number densities $(n_{s_q})$, and Hall parameters $(\beta_{s_q})$. The expressions for the collision frequencies $\nu_m^e$ and $\nu_m^i$ are obtained from [26]. The boundary conditions impose no-slip rotation at both walls, fix the outer pressure to the feed pressure, and prescribe either fixed wall temperatures or an inner adiabatic condition depending on the case studied.

$$\frac{1}{r}\frac{d}{dr}(r \rho_h V_r) = 0$$

$$\frac{d(p_h + p_e)}{dr} = \rho_h \frac{V_\theta^2}{r} + J_\theta B_z$$

$$\frac{1}{r^2}\frac{d}{dr}\left(r^3 \mu_n \frac{d}{dr}\frac{V_\theta}{r}\right) = J_r B_z$$

$$\frac{1}{r}\frac{d}{dr}\left(r\kappa_n \frac{d}{dr} T_h\right) + J_{ri}\left(\frac{J_r - \sigma_{pi}\varepsilon_{ri}^*}{\sigma_p}\right) + \mu_n \left(r \frac{d}{dr}\frac{V_\theta}{r}\right)^2 = \frac{3 n_e k_B m_e \nu_m^e}{m_n}(T_h - T_e)$$

$$\sigma_{pe}\left(\frac{J_r - \sigma_{pi}\varepsilon_{ri}^*}{\sigma_p}\right)^2 = \frac{3 n_e k_B m_e \nu_m^e}{m_n}(T_e - T_h)$$

$$\begin{bmatrix} J_r \\ J_\theta \end{bmatrix} = \frac{n_e q_e^2}{m_e \nu_m^e} \begin{bmatrix} \frac{1}{1+\beta_e^2} & \frac{\beta_e}{1+\beta_e^2} \\ -\frac{\beta_e}{1+\beta_e^2} & \frac{1}{1+\beta_e^2} \end{bmatrix} \begin{bmatrix} E_r^* \\ 0 \end{bmatrix} + \frac{n_i q_i^2}{m_i \nu_m^i} \begin{bmatrix} \frac{1}{1+\beta_i^2} & \frac{\beta_i}{1+\beta_i^2} \\ -\frac{\beta_i}{1+\beta_i^2} & \frac{1}{1+\beta_i^2} \end{bmatrix} \begin{bmatrix} E_r^* + \frac{m_i}{e}\frac{V_\theta^2}{r} \\ 0 \end{bmatrix}$$

$$\frac{n_e^2}{n_n} = 4\left(\frac{2\pi m_e k_B T_e}{h^2}\right)^{3/2} \exp\left(-\frac{\varepsilon_{ion}}{k_B T_e}\right)$$

**Appendix B: Reduced SDC Formulation Implemented in Numerical Model**

The SDC formulation is used as a wall-forced comparison case against the EMDC model, while omitting the higher-fidelity gas-centrifuge flow and transport modeling used in dedicated SDC simulations [44]. It treats the system as a steady, axisymmetric, radially one-dimensional annular SDC working on a gas composed of $\mathcal{S} = \{Ar\}$. The force is imposed through the outer-wall velocity, $\vec{V}(r = R_2) = V_W\hat{\theta}$, while the inner wall is stationary, $\vec{V}(r = R_1) = 0$. The azimuthal velocity profile is therefore determined by viscous transport of angular momentum from the rotating outer boundary through the annulus. The radial pressure profile is then obtained from centrifugal balance, $dp/dr = \rho_n V_\theta^2/r$, with the outer pressure fixed to the feed pressure, $p(r = R_2) = p_{feed}$. Thermal evolution in this reduced SDC model is controlled only by the balance of viscous dissipation ($\mu_n (r d(V_\theta/r)/dr)^2$) and radial heat conduction ($d(r\kappa_n dT/dr)/rdr$), with both walls held at 300 K. This makes the SDC case a simplified shear-forced baseline that isolates how $\alpha_{device}$ scales when momentum is introduced at a rotating boundary rather than volumetrically through $\vec{J} \times \vec{B}$ forcing.

$$\frac{1}{r}\frac{d}{dr}(r\rho_n V_r) = 0$$

$$\frac{dp}{dr} = \rho_n \frac{V_\theta^2}{r}$$

$$\frac{1}{r^2}\frac{d}{dr}\left(r^3 \mu_n \frac{d}{dr}\frac{V_\theta}{r}\right) = 0$$

$$\frac{1}{r}\frac{d}{dr}\left(r\kappa_n \frac{dT}{dr}\right) + \mu_n\left(r\frac{d}{dr}\frac{V_\theta}{r}\right)^2 = 0$$

**References**

[1] Glaser A 2008 Characteristics of the gas centrifuge for uranium enrichment and their relevance for nuclear weapon proliferation Sci. Glob. Secur. 16 1–25

[2] Benedict M, Pigford T H and Levi H W 1981 Nuclear Chemical Engineering (New York: McGraw-Hill)

[3] Kessler G 2011 Proliferation-Proof Uranium and Plutonium Fuel Cycles: Safeguards and Non-Proliferation (Karlsruhe: KIT Scientific Publishing)

[4] Szepessy S and Thorwid P 2018 Low energy consumption of high-speed centrifuges Chem. Eng. Technol. 41 2375–84

[5] Gueroult R, Zweben S J, Fisch N J and Rax J M 2019 E×B configurations for high-throughput plasma mass separation: an outlook on possibilities and challenges Phys. Plasmas 26 043511

[6] Gueroult R and Fisch N J 2014 Plasma mass filtering for separation of actinides from lanthanides Plasma Sources Sci. Technol. 23 035002

[7] U.S. Department of Energy, Office of NEPA Policy and Compliance 2025 CX-270890: Marathon Fusion, Inc. — Partial Ionization Centrifuge Enabling Differential Pumping and Lithium Isotope Separation for Fusion Energy Online: https://www.energy.gov/nepa/articles/cx-270890-marathon-fusion-inc-partial-ionization-centrifuge-enabling-differential (accessed 18 June 2026)

[8] Wijnakker M M B and Granneman E H A 1980 Limitations on mass separation by the weakly ionized plasma centrifuge Z. Naturforsch. A 35 883–93

[9] Wijnakker M M B 1980 Centrifugal effects in a weakly ionized rotating gas PhD Thesis

Universiteit van Amsterdam
[10] Korobtsev S V, Kosinova T A, Rakhimbabaev Y R and Rusanov V D 1986 Chemical processes in a plasma centrifuge Plasma Chem. Plasma Process. 6 97–107
[11] Prasad R R and Krishnan M 1987 Isotope separation in a vacuum-arc centrifuge J. Appl. Phys. 61 4464–70
[12] Egle B, Asgari M, Bigelow T, Bodey I, Burkhardt E, Duckworth R, Goulding R, Robinson S, Tusar J and Mahgerefteh M 2020 Plasma separation process feasibility study for the commercial enrichment of gadolinium-157 Technical Report ORNL/TM-2020/26 (Oak Ridge, TN: Oak Ridge National Laboratory)
[13] Fetterman A J and Fisch N J 2009 Wave-driven countercurrent plasma centrifuge Plasma Sources Sci. Technol. 18 045003
[14] Pronko P P and VanRompay P A 2026 Magnetic centrifuge effects in ultrafast laser ablation plasmas J. Appl. Phys. 139 165902
[15] Wong A Y, Lee K H and Lee L C 2024 Simulation of dynamics of rotating weakly ionized plasmas Phys. Plasmas 31 013101
[16] Lehnert B 1973 The partially ionized plasma centrifuge Phys. Scr. 7 102
[17] Muir H A, Eschbach N, Rodway-Gant G, Vankov I, Chen A, Wrixon B, Li Z, Gunn A and Gregori G 2025 Steady-state rotational dynamics of a weakly ionized hydrogen plasma under cross-field configuration Phys. Plasmas 32 123505
[18] Mitchner M and Kruger C H 1973 Partially Ionized Gases (New York: Wiley)
[19] Inan U S and Gołkowski M 2011 Principles of Plasma Physics for Engineers and Scientists (Cambridge: Cambridge University Press)
[20] LeVan J and Baalrud S D 2025 Foundations of magnetohydrodynamics Phys. Plasmas 32 070901
[21] Meier E T and Shumlak U 2012 A general nonlinear fluid model for reacting plasma-neutral mixtures Phys. Plasmas 19 072508
[22] Raizer Y P 1991 Gas Discharge Physics (Berlin: Springer)
[23] Braginskii S I 1965 Transport processes in a plasma Reviews of Plasma Physics vol 1 ed M A Leontovich (New York: Consultants Bureau) pp 205–311
[24] Kaganovich I D, Smolyakov A, Raitses Y, Ahedo E, Mikellides I G, Jorns B, Taccogna F, Gueroult R, Tsikata S, Bourdon A, Boeuf J-P, Keidar M, Powis A T, Merino M, Cappelli M, Hara K, Carlsson J A, Fisch N J, Chabert P, Schweigert I, Lafleur T, Matyash K, Khrabrov A V, Boswell R W and Fruchtman A 2020 Physics of E×B discharges relevant to plasma propulsion and similar technologies Phys. Plasmas 27 120601
[25] Keidar M and Beilis I I 2006 Electron transport phenomena in plasma devices with E×B drift IEEE Trans. Plasma Sci. 34 804–14
[26] Baeva M, Kozakov R, Gorchakov S and Uhrlandt D 2012 Two-temperature chemically non-equilibrium modelling of transferred arcs Plasma Sources Sci. Technol. 21 055027
[27] Neufeld P D, Janzen A R and Aziz R A 1972 Empirical equations to calculate 16 of the transport collision integrals Ω(l,s)* for the Lennard-Jones (12–6) potential J. Chem. Phys. 57 1100–2
[28] Bird R B, Stewart W E, Lightfoot E and Klingenberg D J 2014 Introductory Transport Phenomena (New York: Wiley)
[29] Dymond J H 1971 High temperature transport coefficients for the rare gases neon to xenon J. Phys. B: At. Mol. Phys. 4 621–7
[30] Fridman A A 2008 Plasma Chemistry (Cambridge: Cambridge University Press)
[31] Hazeltine R D and Waelbroeck F L 2004 The Framework of Plasma Physics (Boulder, CO: Westview Press)
[32] Song P, Gombosi T I and Ridley A J 2001 Three-fluid Ohm's law J. Geophys. Res.: Space Phys. 106 8149–56
[33] Stelzer Z, Cébron D, Miralles S, Vantieghem S, Noir J, Scarfe P and Jackson A 2015 Experimental and numerical study of electrically driven magnetohydrodynamic flow in a modified cylindrical annulus. I. Base flow Phys. Fluids 27 077101
[34] Wilhelm H E and Hong S H 1977 Boundary-value problem for plasma centrifuge at arbitrary magnetic Reynolds numbers Phys. Rev. A 15 2108–16
[35] Underwood T C, Subramaniam V, Riedel W M, Raja L L and Cappelli M A 2019 Effects of

flow collisionality on ELM replication in plasma guns Fusion Eng. Des. 144 97–106
[36] Underwood T C, Loebner K T K, Miller V A and Cappelli M A 2020 Schlieren diagnostic for cinematic visualization of dense plasma jets at Alfvénic timescales Exp. Fluids 61 1
[37] Underwood T C, Loebner K T K, Miller V A and Cappelli M A 2019 Dynamic formation of stable current-driven plasma jets Sci. Rep. 9 2588
[38] Subramaniam V, Underwood T C, Raja L L and Cappelli M A 2018 Computational and experimental investigation of plasma deflagration jets and detonation shocks in coaxial plasma accelerators Plasma Sources Sci. Technol. 27 025016
[39] Bevington P R and Robinson D K 2003 Data Reduction and Error Analysis for the Physical Sciences 3rd edn (New York: McGraw-Hill)
[40] Panjan M, Loquai S, Klemberg-Sapieha J E and Martinu L 2015 Non-uniform plasma distribution in dc magnetron sputtering: origin, shape and structuring of spokes Plasma Sources Sci. Technol. 24 065010
[41] Sengupta M, Smolyakov A and Raitses Y 2021 Restructuring of rotating spokes in response to changes in the radial electric field and the neutral pressure of a cylindrical magnetron plasma J. Appl. Phys. 129 223302
[42] Kaneko O, Sasaki S and Kawashima N 1978 Mass separation experiment with a partially ionized rotating plasma *Plasma Phys.* 20 1167–78.
[43] Lev D and Choueiri E 2012 Scaling of efficiency with applied magnetic field in magnetoplasmadynamic thrusters J. Propuls. Power 28 609–16
[44] Bogovalov S V, Borisevich V D, Borman V D, Kislov V A, Tronin I V and Tronin V N 2013 Verification of numerical codes for modeling of the flow and isotope separation in gas centrifuges Comput. Fluids 86 177–84